\documentclass[pra,aps,groupaddress,showpacs]{revtex4}
\usepackage{graphicx,graphics,epsfig}
\usepackage{dcolumn,array}
\usepackage{bm}
\usepackage{amsmath}
\usepackage{verbatim}
\usepackage{color}
\usepackage{subfigure}
\usepackage{times,natbib}
\usepackage{amsmath,amsfonts,amssymb,graphics,graphicx,epsfig,color,times,natbib}
\usepackage{tikz}
\usepackage{pgfplots}
\usepackage{verbatim}
\usepackage{url}
\usepackage{appendix}
\usepackage[colorlinks=true]{hyperref}

\date{\today}

\begin{document}
\title{$\mathbb{Z}_3$ parafermionic chain emerging from Yang-Baxter equation}

\author{Li-Wei Yu}
\email{NKyulw@yahoo.com}
 \affiliation{Theoretical Physics Division, Chern Institute of Mathematics, Nankai University,
 Tianjin 300071, China}
\author{Mo-Lin Ge}
\email{geml@nankai.edu.cn}
 \affiliation{Theoretical Physics Division, Chern Institute of Mathematics, Nankai University,
 Tianjin 300071, China}



\date{\today}


\begin{abstract}
\textbf{We construct the 1D $\mathbb{Z}_3$ parafermionic model based on the solution of Yang-Baxter equation and express the model by three types of fermions. It is shown that the $\mathbb{Z}_3$ parafermionic chain possesses both triple degenerate ground states and non-trivial topological winding number. Hence, the  $\mathbb{Z}_3$ parafermionic model is a direct generalization of 1D $\mathbb{Z}_2$ Kitaev model.  Both the $\mathbb{Z}_2$ and $\mathbb{Z}_3$ model can be obtained from Yang-Baxter equation.  On the other hand, to show the algebra of parafermionic tripling intuitively, we define a new 3-body Hamiltonian $\hat{H}_{123}$ based on Yang-Baxter equation. Different from the Majorana doubling, the $\hat{H}_{123}$ holds triple degeneracy at each of energy levels. The triple degeneracy is protected by two symmetry operators of the system, $\omega$-parity $P$($\omega=e^{{\textrm{i}\frac{2\pi}{3}}}$) and emergent parafermionic operator $\Gamma$, which are the generalizations of parity $P_{M}$ and emergent Majorana operator in Lee-Wilczek model, respectively. Both the $\mathbb{Z}_3$ parafermionic model and $\hat{H}_{123}$ can be viewed as SU(3) models in color space. In comparison with the Majorana models for SU(2), it turns out that the SU(3) models are truly the generalization of Majorana models resultant from Yang-Baxter equation.} 
\end{abstract}


\maketitle

 The double degeneracy of a pair of Majorana zero modes in condensed matter system has attracted much attentions due to its potential applications in quantum computation and quantum information \cite{kitaev2001unpaired,ivanov2001non,kitaev2006anyons,wilczek2009majorana,leijnse2012introduction,alicea2011non}. It is well known that this topologically protected doubling is immune to local perturbations. Taking 1D p-wave Kitaev model \cite{kitaev2001unpaired} as an example, the Majorana mode appears in the topological phase where the two free Majorana fermions $\gamma_1$ and $\gamma_{2N}$ can be excited without cost of energy at the two ends of the chain model and compose a non-local complex fermion, hence the ground state possesses double degeneracy. The two degenerate states can be differentiated by the electron number parity operator $P_{\textrm{M}}=(-1)^{N_{e}}$, i.e., one state possesses parity $-1$ with odd electron occupation number, while the other possesses parity $+1$ with even electron occupation number.  On the other hand, to give an intuitive analysis about the Majorana doubling, Lee and Wilczek \cite{lee2013algebra} proposed a 3-body Hamiltonian  $\hat{H}_M=\textrm{i}(\alpha \gamma_1\gamma_2+\beta \gamma_2\gamma_3+\kappa \gamma_1\gamma_3)$, where the symmetry operators $P_{\textrm{M}}$ and emergent Majorana operator $\Gamma_{\textrm{M}}$  lead to the doubling at any energy level.

 In our previous paper, we have shown that both the Kitaev model and the Lee-Wilczek model can be derived from the $4\times4$ matrix representation of Yang-Baxter equation(YBE) \cite{yu2015more}. The applications of YBE \cite{yang1967some,baxter1972partition,faddeev1982integrable,kulish1981lecture,korepin1997quantum,takhtajan1989lectures} in constructing many body Hamiltonian had been discussed in various papers\cite{chen2007braiding,chen2008berry,ge2012yang,yu2014factorized}. Specifically, based on the  Majorana representation of Yang-Baxter equation $\breve{R}_\textrm{i}(\theta)=e^{\theta\gamma_\textrm{i}\gamma_\textrm{i+1}}$  , we take the time derivative of $\theta$ in $\breve{R}_\textrm{i}(\theta)$ to obtain the 1D Kitaev model \cite{yu2015more}. To self-contain, we first recall some results related to  $\breve{R}_\textrm{i}(\theta)=e^{\theta\gamma_\textrm{i}\gamma_\textrm{i+1}}$ which emerges from the $4\times4$ matrix representation of YBE, which reads
\begin{equation}\label{4YBEsolution}
\breve{R}'_{\textrm{i}}(\theta)=\left[\begin{array}{cccc}
\cos\theta & 0 & 0 & \sin\theta\\
0 & \cos\theta & \sin\theta & 0\\
0 & -\sin\theta & \cos\theta & 0\\
-\sin\theta & 0 & 0 & \cos\theta
\end{array}\right]=e^{\textrm{i}\theta\sigma_y\otimes\sigma_x}.
\end{equation}
When $\theta=\frac{\pi}{4}$, the $\breve{R}'$-matrix turns into the braid operator $B'$\cite{kauffman2004braiding}
\begin{equation}
B'=\frac{1}{\sqrt{2}}\left[\begin{array}{cccc}
1 & 0 & 0 & 1\\
0 & 1 & 1 & 0\\
0 & -1 & 1 & 0\\
-1 & 0 & 0 & 1
\end{array}\right].
\end{equation}
In our previous paper \cite{yu2015more}, we have shown that the Majorana representation relates to the  $4\times4$ matrix representation of YBE $\breve{R}'_\textrm{i}(\theta)=e^{\textrm{i}\theta\sigma_y^{\textrm{i}}\otimes\sigma_x^{\textrm{i}+1}}$ in tensor product space through Jordan-Wigner(J-W) transformation, here $\sigma_x$ and $\sigma_y$ are Pauli matrices, i and i+1 signify lattice sites. The J-W transformation transforms spin-$\frac{1}{2}$ operators at  lattice sites into spinless fermions through
\begin{equation}\label{JWtrans}
a_{n}^{\dagger}=\left[\prod_{\textrm{i}=1}^{\textrm{n}-1}\sigma^{z}_{\textrm{i}}\right]\sigma^{+}_{n}, \quad  a_{n}=\left[\prod_{\textrm{i}=1}^{\textrm{n}-1}\sigma^{z}_{\textrm{i}}\right]\sigma^{-}_{n},
\end{equation}
where $\sigma^{\pm}_{n}$ are spin ladder operators, $a_{n}^{\dagger}$ and $a_{n}$ are spinless fermions. Define the Majorana fermion\cite{kitaev2001unpaired} 
\begin{equation}\label{MFs}
 \gamma_{2j-1}=a_j^{\dag}+a_j,\, \gamma_{2j}=\textrm{i} (a_j^{\dag}- a_j),\, \{\gamma_i, \gamma_j\}=2\delta_{ij}.
 \end{equation}
Substituting equation (\ref{JWtrans}) and (\ref{MFs}) into equation (\ref{4YBEsolution}),  we can express equation (\ref{4YBEsolution}) as next nearest neighbor interaction of Majorana operators, 
\begin{equation}
\breve{R}'_\textrm{i}(\theta)=e^{\theta\gamma_\textrm{2i-1}\gamma_\textrm{2i+1}},
\end{equation}
where $\gamma_\textrm{2i-1}$ satisfy Clifford algebra $\{\gamma_\textrm{2i-1}, \gamma_\textrm{2j-1}\}=2\delta_\textrm{ij}$. Based on the Clifford algebra, the $\breve{R}'_\textrm{i}(\theta)$ can be rewritten as 
 \begin{equation}
 \breve{R}_\textrm{i}(\theta)=e^{\theta\gamma_\textrm{i}\gamma_\textrm{i+1}}.
\end{equation}
It is easy to check that the $\breve{R}_\textrm{i}(\theta)$ satisfies YBE. Hence the matrix representation of solution of YBE and Majorana representation of braid operators are well related.

A question is raised naturally. One spin site corresponds to 2 subcells of Majorana fermions that are related to $4\times4$ YBE in the tensor space due to the 4-d representation of Temperley-Lieb algebra. On the other hand, the $9\times9$ form of solution of YBE has been known, then could we extend the above discussion to the new type of ``Majorana fermions'' with 3 subcells on one spin site? The answer is yes. Instead of SU(2), the SU(3) operators should naturally be introduced. For the convenience, we call the space color space. 

Since the Majorana models hold 2-fold degeneracy, for SU(3), how can we extend the Majorana double degeneracy to triple degeneracy? In other words, can we construct the extended 1D Kitaev model holding triple degenerate ground state?  Indeed, similar to those in constructing Majorana models via $4\times4$ matrix solution of YBE, we can find the triple degenerate models based on the $9\times9$ matrix representation of YBE \cite{Rowell2009localization}.

In this paper, we make the following progress: 1) Based on the $9\times9$ matrix representation of YBE and the $3\times3$ 3-cyclic representation of SU(3) generators (see Supplementary), we make the decomposition of the $9\times9$ matrix by tensor products of 3-dimensional matrices. By defining generalized SU(3) J-W transformation, we transform the  SU(3) sites into non-local operators and obtain the new representation of YBE. 2) We obtain the $\mathbb{Z}_3$ parafermionic chain with triple degenerate ground states in color space and express the chain with three types of fermions, besides that, the $\mathbb{Z}_N$ case is discussed; 3) In $\mathbb{Z}_3$ parafermionic model, the topological phase transition is signified by the triple degeneracy of ground states and the topological winding number; 4) To give an intuitive explanation of the triple degeneracy and analyse the algebraic structure in it, we construct a 3-body Hamiltonian and find its symmetry operators that lead to the tripling. 

\section*{RESULTS}

\bigskip

\noindent \textbf{Review of two Majorana models.}
To preserve the self-consistency of this paper, firstly, let us give a brief introduction to the construction of Majorana models based on YBE. The intrinsic connection between  the solution $\breve{R}_\textrm{i}(\theta)=e^{\theta\gamma_{i}\gamma_{i+1}}$ of YBE and Kitaev model is that both of them possess $\mathbb{Z}_2$ symmetry. Next we review the Kitaev model derived from YBE. 
 
We imagine that a unitary evolution is governed by $\breve{R}_{\textrm{i}}(\theta)$. If only $\theta$ ($\tan\theta$ is the velocity $u$ of a particle) in unitary operator $\breve{R}_{\textrm{i}}(\theta)$ is time-dependent, we can express a state $|\psi(t)\rangle$ as $|\psi(t)\rangle=\breve{R}_{\textrm{i}}(\theta(t))|\psi(0)\rangle$. Taking the Schr\"{o}dinger equation $\textrm{i}\hbar\tfrac{\partial}{\partial t}|\psi(t)\rangle=\hat{H}(t)|\psi(t)\rangle$ into account, one obtains:
\begin{equation}
\textrm{i}\hbar\tfrac{\partial}{\partial t}[\breve{R}_\textrm{i}|\psi(0)\rangle]=\hat{H}(t)\breve{R}_\textrm{i}|\psi(0)\rangle.
\end{equation}
Then the Hamiltonian $\hat{H}_\textrm{i}(t)$ related to the unitary operator $\breve{R}_{\textrm{i}}(\theta)$ is given by:
\begin{equation}\label{SchrodingerEquation}
\hat{H}_\textrm{i}(t)=\textrm{i}\hbar\tfrac{\partial\breve{R}_{\textrm{i}}}{\partial t}\breve{R}_{\textrm{i}}^{-1}.
\end{equation}
Substituting $\breve{R}_\textrm{i}(\theta)=e^{\theta\gamma_{i}\gamma_{i+1}}$ into equation (\ref{SchrodingerEquation}), we have
\begin{equation}\label{2MFHamiltonian}
\hat{H}_\textrm{i}(t)=\textrm{i}\hbar\dot{\theta}\gamma_{i}\gamma_{i+1}.
\end{equation}
If we only consider the nearest-neighbour interactions between two Majorana fermions(MFs) and extend equation (\ref{2MFHamiltonian}) to an inhomogeneous chain with 2N sites, the derived  chain model is expressed as \cite{yu2015more}:
\begin{equation}\label{YBEKitaev}
\hat{H}_\textrm{K}=\textrm{i}\hbar\sum_{k=1}^{N}(\dot{\theta}_1\gamma_{2k-1}\gamma_{2k}+\dot{\theta}_2\gamma_{2k}\gamma_{2k+1}),
\end{equation}
with $\dot{\theta}_1$  and $\dot{\theta}_2$ describing odd-even and even-odd pairs, respectively. This is exactly the Kitaev model derived from YBE.

The properties of 1D Kitaev model  are well known:
\begin{enumerate}
\item In the case $\dot{\theta}_1>0$, $\dot{\theta}_2=0$,  the Hamiltonian reads:
\begin{equation}\label{YBEtrivil}
\hat{H}_{\textrm{triv}}=\textrm{i}\hbar\sum_{k}^{N}\dot{\theta}_1\gamma_{2k-1}\gamma_{2k}.
\end{equation}
As defined in equation (\ref{MFs}), the Majorana operators $\gamma_{2k-1}$ and $\gamma_{2k}$ come from the same ordinary fermion site k, $\textrm{i}\gamma_{2k-1}\gamma_{2k}=2a_{k}^{\dag}a_{k}-1$ ($a_{k}^{\dag}$ and $a_{k}$ are spinless ordinary fermion operators). $\hat{H}_1$ simply means the total occupancy of ordinary fermions in the chain and has U(1) symmetry, $a_j\rightarrow e^{\textrm{i}\phi}a_j $. The ground state represents the ordinary fermion occupation number 0. This Hamiltonian corresponds to the trivial case of Kitaev's. 
\item  In the case $\dot{\theta}_1=0$, $\dot{\theta}_2>0$, the Hamiltonian reads:
\begin{equation}\label{YBEtopo}
\hat{H}_{\textrm{topo}}=\textrm{i}\hbar\sum_{k}^{N}\dot{\theta}_2\gamma_{2k}\gamma_{2k+1}.
\end{equation}
This Hamiltonian corresponds to the topological phase of 1D Kitaev model and has $\mathbb{Z}_2$ symmetry, $a_j\rightarrow -a_j$. Here the operators $\gamma_1$ and $\gamma_{2N}$ are absent in $\hat{H}_2$. The Hamiltonian has two degenerate ground state, $|0\rangle$ and $|1\rangle=d^{\dag}|0\rangle$, $d^{\dag}=(\gamma_1-\textrm{i}\gamma_{2N})/2$. This mode is the so-called Majorana mode in 1D Kitaev model. 
\end{enumerate}

On the other hand, as pointed out by Lee and Wilczek in Ref. \cite{lee2013algebra}, the double degeneracy of Majorana models $\hat{H}_\textrm{M}$ is due to two symmetry operators, the parity operator $P_\textrm{M}$ and emergent Majorana operator $\Gamma_\textrm{M}$. For instance, in 3-body Majorana model with the Hamiltonian
\begin{equation}
\hat{H}_\textrm{M}=\textrm{i}\left(\alpha \gamma_1\gamma_2+\beta\gamma_2\gamma_3+\kappa\gamma_1\gamma_3\right),
\end{equation}
the symmetry operators and commutation relations are
\begin{eqnarray}
&&P_\textrm{M}=(-1)^{N_{e}}, \, \Gamma_\textrm{M}=-\textrm{i}\gamma_1\gamma_2\gamma_3;\\
&&[\hat{H}_\textrm{M}, P_\textrm{M}]=0, \, [\hat{H}_\textrm{M}, \Gamma_\textrm{M}]=0, \, \{\Gamma_\textrm{M}, P_\textrm{M}\}=0.
\end{eqnarray} 
Clearly, in the basis of both $\hat{H}_\textrm{M}$ and $P_\textrm{M}$ are diagonal, $\Gamma_\textrm{M}$ transforms the states with $P_\textrm{M}=\pm1$ into the states with $P_\textrm{M}=\mp1$. Therefore the Hamiltonian possesses Majorana doubling.

\bigskip

\noindent \textbf{Yang-Baxter equation and 3-cyclic SU(3) generators.}
Since the YBE is properly applied in constructing $\mathbb{Z}_2$ Kitaev model,  we try to extend the result to $\mathbb{Z}_3$-symmetric model. Fortunately, the known $9\times9$ matrix representation of the solution to YBE is a proper unitary operator for constructing the desired model with $\mathbb{Z}_3$  symmetry.  Now we give a brief introduction to the $9\times9$ matrix representation of the solution to YBE which is associated with this paper. Firstly, let us introduce the braid matrix \cite{Rowell2009localization} for $\omega=e^{{\textrm{i}\frac{2\pi}{3}}}$: 
\begin{equation}
B=\frac{\textrm{i}}{\sqrt{3}}\left[\begin{array}{ccccccccc}
\omega & 0 & 0 & 0 & 0 & 1 & 0 & \omega & 0\\
0 & \omega & 0 & \omega^2 & 0 & 0 & 0 & 0 & \omega^2\\
0 & 0 & \omega & 0 & \omega & 0 & 1 & 0 & 0\\
0 & \omega^2 & 0 & \omega & 0 & 0 & 0 & 0 & \omega^2\\ 
0& 0 & 1 & 0 & \omega & 0 & \omega & 0 & 0\\
\omega & 0 & 0 & 0 & 0 & \omega & 0 & 1 & 0\\
0 & 0 & \omega & 0 & 1 & 0 & \omega & 0 & 0\\
1 & 0 & 0 & 0 & 0 & \omega & 0 & \omega & 0\\
0 & \omega^2 & 0 & \omega^2 & 0 & 0 & 0 & 0 & \omega 
\end{array}\right],
\end{equation}
which satisfies the braid relation
\begin{equation}
B_{\textrm{i}}B_{\textrm{i+1}}B_{\textrm{i}}=B_{\textrm{i+1}}B_{\textrm{i}}B_{\textrm{i+1}},
\end{equation}
where $B_{\textrm{i}}=I\otimes...I\otimes\mathop{B}\limits_{\textrm{i,i+1}}\otimes I...$ ($I$ is $3\times3$ identity matrix).

The solution $\breve{R}_{\textrm{i}}(\theta)$ of Yang-Baxter equation can be viewed as the parametrization of braid operators, 
\begin{equation}\label{su3Solution}
\begin{split}
\breve{R}_{\textrm{i}}(\theta)=&e^{\textrm{i}\theta M}=\cos\theta+\textrm{i}\sin\theta M,\\
M=&\tfrac{2}{\sqrt{3}}T_{\textrm{i}}-1, \quad M^2=1,
\end{split}
\end{equation}
where $T_\textrm{i}=e^{\textrm{i}\frac{\pi}{6}}(I+B_{\textrm{i}})$ satisfies the Temperley-Lieb algebraic (TLA) relation\cite{Temperley1971relations}
\begin{equation}\label{TLA}
\begin{split}
&T_{\textrm{i}}^2=dT_{\textrm{i}},\quad d=\sqrt{3}\\
&T_{\textrm{i}}T_{\textrm{i}\pm1}T_{\textrm{i}}=T_{\textrm{i}},\\
&T_{\textrm{i}}T_{\textrm{j}}=T_{\textrm{j}}T_{\textrm{i}}, \quad |i-j|>1.
\end{split}
\end{equation}
Then the YBE\cite{jimbobook} means that a 3-body S-matrix can be expressed in terms of three 2-body S-matrices, i.e.:
\begin{equation}\label{YBEtheta}
\breve{R}_{\textrm{i}}(\theta_1) \breve{R}_{\textrm{i}+1}(\theta_2) \breve{R}_{\textrm{i}}(\theta_3)=\breve{R}_{\textrm{i}+1}(\theta_3) \breve{R}_{\textrm{i}}(\theta_2) \breve{R}_{\textrm{i}+1}(\theta_1).
\end{equation}
Substituting equation (\ref{su3Solution}) into equation (\ref{YBEtheta}), we have the constraint for three parameters $\theta_1$, $\theta_2$ and $\theta_3$ :
\begin{equation}\label{YBEangularrelation}
\tan\theta_2=\frac{\tan\theta_1+\tan\theta_3}{1+\frac{1}{3}\tan\theta_1\tan\theta_3}.
\end{equation}
When $\theta_1=\theta_2=\theta_3=\frac{\pi}{3}$, the YBE turns into the braid relation with $\breve{R}_{\textrm{i}}(\pi/3)=\omega B_{\textrm{i}}$. Note that due to the different $d$ in TLA for $4\times4$ and $9\times9$ solutions of YBE($d=\sqrt{2}$ for $4\times4$), the angular relation in equation (\ref{YBEangularrelation}) for $9\times9$  is different from the angular relation for $4\times4$ in equation (\ref{4YBEsolution}), 
\begin{equation}\label{4YBEangularrelation}
\tan\theta'_2=\frac{\tan\theta'_1+\tan\theta'_3}{1+\tan\theta'_1\tan\theta'_3}.
\end{equation}
It is well known that the physical meaning of $\tan\theta_\textrm{i}$(or $\tan\theta'_\textrm{i}$) for $i=1,2,3$ is the velocity of a particle. The angular relation for $4\times4$ solution means the Lorentz addition of the velocity.
 
By introducing the 3-cyclic SU(3) generators $T_{\textrm{i}}^{\textrm{(j)}}$ based on the principal representation of V. Kac\cite{Kac1994infinite} (see Supplementary), TLA generator can be expressed as the tensor product of nearest SU(3)-lattice sites,  
\begin{equation}\label{SU3Tgeneartor}
T_{\textrm{i}}=\frac{1}{\sqrt{3}}(I^{\otimes2}+\omega T_{3}^{(2)}\otimes T_{3}^{(3)}+\omega^2 T_{2}^{(3)}\otimes T_{2}^{(2)})_{\textrm{i,i+1}}.
\end{equation}
Here \cite{liu2013generalized} 
\begin{equation}
\begin{split}
&T_{2}^{(2)}=\left[\begin{array}{ccc} 0 & 1 & 0 \\ 0 & 0 & \omega \\ \omega^2 & 0 & 0\end{array}\right], T_{2}^{(3)}=\left[\begin{array}{ccc} 0 & 0 & 1 \\ \omega & 0 & 0 \\ 0 & \omega^2 & 0\end{array}\right];\\
&T_{3}^{(2)}=\left[\begin{array}{ccc} 0 & 1 & 0 \\ 0 & 0 & \omega^2 \\ \omega & 0 & 0\end{array}\right], T_{3}^{(3)}=\left[\begin{array}{ccc} 0 & 0 & 1 \\ \omega^2 & 0 & 0 \\ 0 & \omega & 0\end{array}\right].
\end{split}
\end{equation}
Hence at each lattice site there is one SU(3) operator which can be identified with the colors blue, red and green \cite{Greite2007exact}.
In the following sections, we will make use of equation (\ref{SU3Tgeneartor}) to generate the topological non-trivial models and triple degeneracy.

\bigskip

\noindent \textbf{Ladder operators of SU(3) spin and extended Jordan-Wigner transformation.}
In this section, we present the ladder operators of SU(3) spin and introduce the extended Jordan-Wigner transformation for  SU(3) spin sites. 

For spin-$\frac{1}{2}$ at lattice sites expressed by SU(2) Pauli matrices, the ladder operators are 
\begin{equation*}
\sigma^{+}=\left[\begin{array}{cc} 0 & 1 \\ 0 & 0 \end{array}\right], \, \sigma^{-}=\left[\begin{array}{cc} 0 & 0 \\1 & 0 \end{array}\right].
\end{equation*}
Similarly, we introduce the cyclic ladder operators of SU(3) spin
\begin{equation}\label{su3ladder}
\begin{split}
u^{+}=\left[\begin{array}{ccc} 0 & 1 & 0 \\ 0 & 0 & 0 \\ 0 & 0 & 0\end{array}\right], s^{+}=\left[\begin{array}{ccc} 0 & 0 & 0 \\ 0 & 0 & \omega \\ 0 & 0 & 0 \end{array}\right], d^{+}=\left[\begin{array}{ccc} 0 & 0 & 0 \\ 0 & 0 & 0 \\ \omega^2 & 0 & 0 \end{array}\right],\\
u^{-}=\left[\begin{array}{ccc} 0 & 0 & 0 \\ 1 & 0 & 0 \\ 0 & 0 & 0\end{array}\right], s^{-}=\left[\begin{array}{ccc} 0 & 0 & 0 \\ 0 & 0 & 0 \\ 0 & \omega^2 & 0 \end{array}\right], d^{-}=\left[\begin{array}{ccc} 0 & 0 & \omega \\ 0 & 0 & 0 \\ 0 & 0 & 0 \end{array}\right].
\end{split}
\end{equation}
The above operators act on the color space to transform the three colors into each other obeying the algebraic relations
\begin{equation}\label{usdrelation}
\begin{split}
&[u^+]^2=[s^+]^2=[d^+]^2=0,\\
&[u^-]^2=[s^-]^2=[d^-]^2=0,\\
&u^{+}s^{+}=d^{-}, \, s^{+}d^{+}=u^{-}, \, d^{+}u^{+}=s^{-},\\
&s^{-}u^{-}=d^{+}, \, d^{-}s^{-}=u^{+}, \, u^{-}d^{-}=s^{+},\\
&u^{+}s^{+}d^{+}+s^{+}d^{+}u^{+}+d^{+}u^{+}s^{+}=1.
\end{split}
\end{equation}
Unlike SU(2) spin, the SU(3) spin has 3 independent ladder operators $u^+$, $s^+$ and $d^+$ to make the transition between three different colors. $u$, $s$ and $d$ can be expressed by $u^+$, $s^+$ and $d^+$.

To introduce the extended Jordan-Wigner transformation for SU(3), let us review the original SU(2) J-W transformation firstly.  J-W transformation transforms sited spin-$\frac{1}{2}$ operators onto spinless fermions
\begin{equation}
a_{n}^{\dagger}=\left[\prod_{\textrm{i}=1}^{\textrm{n}-1}\sigma^{z}_{\textrm{i}}\right]\sigma^{+}_{n}, \quad  a_{n}=\left[\prod_{\textrm{i}=1}^{\textrm{n}-1}\sigma^{z}_{\textrm{i}}\right]\sigma^{-}_{n},
\end{equation}
where $a_{m}^{\dagger}$ and $a_{m}$ satisfy the fermionic commutation relations
\begin{equation}
\{a_{m}^{\dagger},a_{n}\}=\delta_{mn}, \, \{a_{m}^{\dagger},a_{n}^{\dagger}\}=\{a_{m},a_{n}\}=0.
\end{equation}
From the above transformation, it turns out that the anti commuting relations of fermions result from $\{\sigma^{z}_{\textrm{i}}, \sigma^{+}_{\textrm{i}}\}=\{\sigma^{z}_{\textrm{i}}, \sigma^{-}_{\textrm{i}}\}=0$. 

Similarly, the extended J-W transformation for SU(3) can be defined as\cite{fendley2012parafermionic}
\begin{equation}\label{USD}
\begin{split}
&U_{\textrm{n}}^{\dagger}=\left[\prod_{\textrm{i}=1}^{\textrm{n}-1}[T_{2}^{(1)}]_{\textrm{i}}\right]u^{+}_{n},\quad U_{\textrm{n}}=\left[\prod_{\textrm{i}=1}^{\textrm{n}-1}[T_{3}^{(1)}]_{\textrm{i}}\right]u^{-}_{n};\\
&S_{\textrm{n}}^{\dagger}=\left[\prod_{\textrm{i}=1}^{\textrm{n}-1}[T_{2}^{(1)}]_{\textrm{i}}\right]s^{+}_{n},\quad S_{\textrm{n}}=\left[\prod_{\textrm{i}=1}^{\textrm{n}-1}[T_{3}^{(1)}]_{\textrm{i}}\right]s^{-}_{n};\\
&D_{\textrm{n}}^{\dagger}=\left[\prod_{\textrm{i}=1}^{\textrm{n}-1}[T_{2}^{(1)}]_{\textrm{i}}\right]d^{+}_{n}, \quad D_{\textrm{n}}=\left[\prod_{\textrm{i}=1}^{\textrm{n}-1}[T_{3}^{(1)}]_{\textrm{i}}\right]d^{-}_{n},
\end{split}
\end{equation}
where\cite{liu2013generalized} 
\begin{equation*}
T_{2}^{(1)}=\left[\begin{array}{ccc} 1 & 0 & 0 \\ 0 & \omega & 0 \\ 0 & 0 & \omega^2 \end{array}\right], \, T_{3}^{(1)}=\left[\begin{array}{ccc} 1 & 0 & 0 \\ 0 & \omega^2 & 0 \\ 0 & 0 & \omega \end{array}\right],
\end{equation*}
in which it follows
\begin{eqnarray}
&&X_{\textrm{i}}Y_{\textrm{j}}=\omega Y_{\textrm{j}}X_{\textrm{i}} \quad \left( \{X, Y\}\in \{U, S, D\}; \, \textrm{i}<\textrm{j} \right);\\
&&S_{\textrm{i}}U_{\textrm{i}}=D_{\textrm{i}}^{\dagger}, \, D_{\textrm{i}} S_{\textrm{i}}=U_{\textrm{i}}^{\dagger}, \, U_{\textrm{i}} D_{\textrm{i}}=S_{\textrm{i}}^{\dagger},\label{USDrelations}
\end{eqnarray}
that can be checked straightforwardly. Different from the anti commuting of spinless fermions, in the exchange between the above operators there appear extra $\omega$ (or $\omega^2$) phase factor. The physical meaning is obvious, when making exchange between two particles on $i$-th and $j$-th sites ($i<j$), the system gains an extra $\omega$ phase factor. Exchanging the two particles again, the system returns to the initial state. 

By introducing the linear combination of sited SU(3) operators
\begin{eqnarray}
&&F_{\textrm{i}}^{\dagger}=U_{\textrm{i}}^{\dagger}+\omega^2S_{\textrm{i}}^{\dagger}+\omega D_{\textrm{i}}^{\dagger}, \quad G_{\textrm{i}}^{\dagger}=U_{\textrm{i}}^{\dagger}+S_{\textrm{i}}^{\dagger}+ D_{\textrm{i}}^{\dagger};\label{fgd}\\
&&F_{\textrm{i}}=U_{\textrm{i}}+\omega S_{\textrm{i}}+\omega^2 D_{\textrm{i}}, \qquad G_{\textrm{i}}=U_{\textrm{i}}+ S_{\textrm{i}}+ D_{\textrm{i}},\label{fg}\\
&&F_{\textrm{i}}^{\dagger}=F_{\textrm{i}}^2, \, G_{\textrm{i}}^{\dagger}=G_{\textrm{i}}^2, \, F_{\textrm{i}}^3=G_{\textrm{i}}^3=1;\\
&&X_{\textrm{i}}Y_{\textrm{j}}=\omega Y_{\textrm{j}}X_{\textrm{i}} \quad \left( \{X, Y\}\in \{F, G \}; \, \textrm{i}<\textrm{j} \right),
\end{eqnarray}
the TLA generator $T_{\textrm{i}}$ in equation (\ref{SU3Tgeneartor}) can be rewritten as
\begin{equation}\label{FGTLA}
T_{\textrm{i}}=\frac{1}{\sqrt{3}}\left[1+\omega F_{\textrm{i}}^{\dagger}G_{\textrm{i}+1}+F_{\textrm{i}}G_{\textrm{i}+1}^{\dagger}\right].
\end{equation}
Here  $F_{\textrm{i}}^{\dagger}$ and $G_{\textrm{i}}^{\dagger}$ are non-local operators. Indeed, they are the generalizations of Clifford algebra which corresponds to Majorana fermions. Redefining
\begin{eqnarray}
&&C^{\dagger}_{2\textrm{i}-1}=F_{\textrm{i}}^{\dagger}, \quad C_{2\textrm{i}-1}=F_{\textrm{i}};\label{cf}\\
&&C^{\dagger}_{2\textrm{i}}=\omega G_{\textrm{i}}^{\dagger}, \quad  C_{2\textrm{i}}=\omega^{2}G_{\textrm{i}}\label{cg}.
\end{eqnarray}
Thus the extended generator of TLA shown in equation (\ref{FGTLA}) can be written in terms of the form
\begin{equation}\label{Crepresentation}
T_{\textrm{i}}'=\frac{1}{\sqrt{3}}\left[1+\omega^2 C_{\textrm{i}}^{\dagger}C_{\textrm{i}+1}+\omega^2C_{\textrm{i}}C_{\textrm{i}+1}^{\dagger}\right],
\end{equation}
with
\begin{eqnarray}
&&C_{\textrm{i}} C_{\textrm{j}}=\omega C_{\textrm{j}} C_{\textrm{i}}, \quad \left(i<j\right),\label{OmegaCommute}\\
&&C_{\textrm{i}}=[C_{\textrm{i}}^{\dagger}]^2, \quad[C_{\textrm{i}}]^3=[C_{\textrm{i}}^{\dagger}]^3=1.
\end{eqnarray}
For convenience we call the commutation relation shown in equation (\ref{OmegaCommute}) $\omega$-commutation relation. This commutation relation can be regarded as the generalization of Majorana fermions' anti-commuting, which is also proposed in \cite{cobanera2014fock}. Note that $T_{\textrm{i}}'$ does not equal to $T_{\textrm{i}}$, but it  also satisfies $d=\sqrt{3}$ TLA in equation (\ref{TLA}) and can be substituted into equation (\ref{su3Solution}).  In 1D Kitaev model, two real Majorana operators corresponds to one complex fermion site as well as one SU(2) spin site. Similarly, the two $\omega$-commuting operators $C_{\textrm{2i}-1}$ and $C_{\textrm{2i}}$ correspond to the $i$-th SU(3) spin.   Obviously, equation (\ref{OmegaCommute}) looks $q$-commutation relation for $q^3=1$ in quantum algebra \cite{wiegmann1994bethe}.

\bigskip

\noindent \textbf{Generating $\mathbb{Z}_3$ parafermionic model from YBE.}
From equation (\ref{su3Solution}) and equation (\ref{Crepresentation}), we obtain the unitary solution $\breve{R}_{\textrm{i}}(\theta)$ of YBE in the form
\begin{equation}\label{CrepR}
\breve{R}_{\textrm{i}}(\theta)=e^{-\textrm{i}\theta}+\textrm{i}\frac{2}{3}\sin\theta\left[1+\omega^2C_{\textrm{i}}^{\dagger}C_{\textrm{i}+1}+\omega^2C_{\textrm{i}}C_{\textrm{i}+1}^{\dagger}\right].
\end{equation}
Now let us construct the $\mathbb{Z}_3$ parafermionic chain based on equation (\ref{CrepR}). Substituting equation (\ref{CrepR}) into equation (\ref{SchrodingerEquation}), we get
\begin{equation}\label{2CHamiltonian}
\hat{H}_{\textrm{i}}(t)=-\dot{\theta}\frac{2\hbar}{3}\left(\omega^2C_{\textrm{i}}^{\dagger}C_{\textrm{i}+1}+\omega^2C_{\textrm{i}}C_{\textrm{i}+1}^{\dagger}-\frac{1}{2}\right).
\end{equation}
Similarly, we consider the nearest-neighbour interactions of $C_{\textrm{i}}$'s and extend equation (\ref{2CHamiltonian}) to an 2N-chain  and ignore the constant term, the derived chain model can be expressed as:
\begin{equation}\label{YBEcolorKitaev}
\hat{H}=-\frac{2\hbar}{3}\omega^2{\Big [ }\sum_{i=1}^{N}\dot{\theta}_1\left(C_{\textrm{2i}-1}^{\dagger}C_{\textrm{2i}}+C_{\textrm{2i}-1}C_{\textrm{2i}}^{\dagger}\right)+\sum_{i=1}^{N-1}\dot{\theta}_2\left(C_{\textrm{2i}}^{\dagger}C_{\textrm{2i}+1}+C_{\textrm{2i}}C_{\textrm{2i}+1}^{\dagger}\right)\Big].
\end{equation}
 Here we emphasize that the chain possesses open boundary condition. This model is the 1D $\mathbb{Z}_3$ parafermionic model\cite{fendley2012parafermionic}, which originates from the three-state Potts model\cite{stephen1972pseudo,fateev1982self,baxter1988new}. Instead of the $\mathbb{Z}_2$ parity symmetry of Kitaev model, the model in equation (\ref{YBEcolorKitaev}) possesses $\mathbb{Z}_3$ symmetry.  The symmetry operator is 
\begin{equation}
P=\prod_{\textrm{i}=1}^{\textrm{N}}\left(C_{2\textrm{i}-1}^{\dagger}C_{2\textrm{i}}\right), \quad P^3=1.
\end{equation}
Hence $P$ is a $\mathbb{Z}_3$ symmetry of the model and the eigenvalues of $P$ is $1$, $\omega$ and $\omega^2$. Next we analyse the obtained model in two cases.
\begin{enumerate}
\item $\dot{\theta}_1>0$, $\dot{\theta}_2=0.$

In this case the Hamiltonian becomes:
\begin{eqnarray}
\hat{H}_1&=&-\frac{2\hbar}{3}\dot{\theta}_1\omega^2\sum_{i=1}^{N}\left(C_{\textrm{2i}-1}^{\dagger}C_{\textrm{2i}}+C_{\textrm{2i}-1}C_{\textrm{2i}}^{\dagger}\right)\nonumber\\
&=&\frac{2\hbar}{3}\dot{\theta}_1\sum_{i=1}^{N}\left[d_{\textrm{i}}^{\dagger}d_{\textrm{i}}-2\right].\label{YBESU3trivil}
\end{eqnarray}

\begin{widetext}
\newsavebox\tensorsitetwo
\begin{lrbox}{\tensorsitetwo}
  \begin{minipage}{0.3\textwidth}
     $${\scriptstyle \sigma_{\textrm{k}}^{+}, \, \sigma_{\textrm{k}}^{-}} $$ 
  \end{minipage}
\end{lrbox}

\newsavebox\tensorsitethree
\begin{lrbox}{\tensorsitethree}
  \begin{minipage}{0.3\textwidth}
     \[ {\scriptstyle u_{\textrm{k}}^{+}, \, s_{\textrm{k}}^{+}, \, d_{\textrm{k}}^{+} }\]
  \end{minipage}
\end{lrbox}

\newsavebox\jwtranstwo
\begin{lrbox}{\jwtranstwo}
  \begin{minipage}{0.3\textwidth}
    \begin{align*}
     {\scriptstyle a_{\textrm{k}}^{\dag}, \, a_{\textrm{k}}}
    \end{align*} 
  \end{minipage}
\end{lrbox}

\newsavebox\jwtransthree
\begin{lrbox}{\jwtransthree}
  \begin{minipage}{0.3\textwidth}
    \begin{align*}
     {\scriptstyle U_{\textrm{k}}^{\dag}, \, S_{\textrm{k}}^{\dag}, \, D_{\textrm{k}}^{\dag}}
    \end{align*} 
  \end{minipage}
\end{lrbox}

\newsavebox\qptwo
\begin{lrbox}{\qptwo}
  \begin{minipage}{0.3\textwidth}
    \begin{align*}
     &\quad{\scriptstyle \gamma_{\textrm{2k-1}}=a_{\textrm{k}}^{\dag}+a_{\textrm{k}};} \\
     &\quad {\scriptstyle \gamma_{\textrm{2k}}=\textrm{i}(a_{\textrm{k}}^{\dag}-a_{\textrm{k}}).}\\
     &\, \, {\scriptstyle \textrm{(Majorana fermions)}}
    \end{align*} 
  \end{minipage}
\end{lrbox}

\newsavebox\qpthree
\begin{lrbox}{\qpthree}
  \begin{minipage}{0.3\textwidth}
    \begin{align*}
     &{\scriptstyle C_{\textrm{2k-1}}^{\dag}=U_{\textrm{k}}^{\dag}+ \omega^2S_{\textrm{k}}^{\dag}+ \omega D_{\textrm{k}}^{\dag}; }\\  
     &{\scriptstyle C_{\textrm{2k}}^{\dag}=\omega U_{\textrm{k}}^{\dag}+ \omega S_{\textrm{k}}^{\dag}+ \omega D_{\textrm{k}}^{\dag}}.
    \end{align*} 
  \end{minipage}
\end{lrbox}

\newsavebox\crtwo
\begin{lrbox}{\crtwo}
  \begin{minipage}{0.3\textwidth}
    \begin{align*}
   &{\scriptstyle \gamma_{\textrm{i}}\gamma_{\textrm{j}}=-\gamma_{\textrm{j}}\gamma_{\textrm{i}};}\\
    &{\scriptstyle \gamma_{\textrm{i}}^2=1, \, \gamma_{\textrm{i}}=\gamma_{\textrm{i}}^{\dag}.}
         \end{align*} 
  \end{minipage}
\end{lrbox}

\newsavebox\crthree
\begin{lrbox}{\crthree}
  \begin{minipage}{0.3\textwidth}
    \begin{align*}
     &{\scriptstyle C_{\textrm{i}}^{\dag}C_{\textrm{j}}^{\dag}=\omega C_{\textrm{j}}^{\dag}C_{\textrm{i}}^{\dag}, \, (\textrm{j}>\textrm{i});} \\
     &{\scriptstyle [C_{\textrm{i}}^{\dag}]^3=1, \, [C_{\textrm{i}}]^2=C_{\textrm{i}}^{\dag}.}
    \end{align*} 
  \end{minipage}
\end{lrbox}

\newsavebox\botwo
\begin{lrbox}{\botwo}
  \begin{minipage}{0.3\textwidth}
    \begin{align*}
   &{\scriptstyle B_{\textrm{i}}'=\frac{1}{\sqrt{2}}\left(1+\gamma_{\textrm{i}}\gamma_{\textrm{i}+1}\right);}\\
    &\qquad \,{\scriptstyle [B_{\textrm{i}}']^8=1.}
         \end{align*} 
  \end{minipage}
\end{lrbox}

\newsavebox\bothree
\begin{lrbox}{\bothree}
  \begin{minipage}{0.3\textwidth}
    \begin{align*}
     &{\scriptstyle B_{\textrm{i}}=\frac{\textrm{i}}{\sqrt{3}}\omega^2\left(1+C_{\textrm{i}}^{\dag}C_{\textrm{i}+1}+C_{\textrm{i}}C_{\textrm{i}+1}^{\dag}\right);}\\
    & \qquad \qquad \, {\scriptstyle [B_{\textrm{i}}]^6=1.}
    \end{align*} 
  \end{minipage}
\end{lrbox}

\newsavebox\cotwo
\begin{lrbox}{\cotwo}
  \begin{minipage}{0.3\textwidth}
    \begin{align*}
   &{\scriptstyle B_{\textrm{i}}'^{\dag}\gamma_{\textrm{i}}B_{\textrm{i}}'=\gamma_{\textrm{i}+1};}\\
  &{\scriptstyle B_{\textrm{i}}'^{\dag}\gamma_{\textrm{i}+1}B_{\textrm{i}}'=-\gamma_{\textrm{i}};}\\
  &{\scriptstyle B_{\textrm{i}}'^{\dag}(-\gamma_{\textrm{i}})B_{\textrm{i}}'=-\gamma_{\textrm{i}+1};}\\
  &{\scriptstyle B_{\textrm{i}}'^{\dag}(-\gamma_{\textrm{i}+1})B_{\textrm{i}}'=\gamma_{\textrm{i}}.}
           \end{align*} 
  \end{minipage}
\end{lrbox}

\newsavebox\cothree
\begin{lrbox}{\cothree}
  \begin{minipage}{0.3\textwidth}
    \begin{align*}
     &{\scriptstyle B_{\textrm{i}}^{\dag}C_{\textrm{i}}^{\dag}B_{\textrm{i}}=C_{\textrm{i}+1}^{\dag};}\\
  &{\scriptstyle B_{\textrm{i}}^{\dag}C_{\textrm{i}+1}^{\dag}B_{\textrm{i}}=\omega^2 C_{\textrm{i}+1}C_{\textrm{i}};}\\
  &{\scriptstyle B_{\textrm{i}}^{\dag}(\omega^2 C_{\textrm{i}+1}C_{\textrm{i}})B_{\textrm{i}}=C_{\textrm{i}}^{\dag}.} 
     \end{align*} 
  \end{minipage}
\end{lrbox}

\begin{table}
  \caption{\bf Comparison between SU(2) and SU(3) pictures \label{tab:Compare23}}
  \begin{center}
    \begin{tabular}{lcc}
    \hline\hline  \smallskip
      Type of spin sites & SU(2) & SU(3) \\ \hline \smallskip\\
      Ladder operators &   ${\scriptstyle \sigma_{\textrm{k}}^{+}, \, \sigma_{\textrm{k}}^{-}.}$ & ${\scriptstyle u_{\textrm{k}}^{+}, \, s_{\textrm{k}}^{+}, \, d_{\textrm{k}}^{+}.}$ \medskip\\ 
      After J-W transform.  & ${\scriptstyle a_{\textrm{k}}^{\dag}, \, a_{\textrm{k}}.}$ &  ${\scriptstyle U_{\textrm{k}}^{\dag}, \, S_{\textrm{k}}^{\dag}, \, D_{\textrm{k}}^{\dag}.}$ \\\smallskip
      Quasiparticle & \usebox{\qptwo} & \usebox{\qpthree}\\
      Commutation relation & \usebox{\crtwo} & \usebox{\crthree}\\
      Braid operator & \usebox{\botwo} & \usebox{\bothree}\\
      Cyclic operation & \usebox{\cotwo} & \usebox{\cothree}\\\hline\hline
    \end{tabular}
  \end{center}
\end{table}
\end{widetext}
Here we note that $C_{\textrm{2i}-1}$ and $C_{\textrm{2i}}$ correspond to i-th SU(3) spin, $d_{\textrm{i}}^{\dagger}=C_{\textrm{2i-1}}^{\dag}-\omega C_{\textrm{2i}}^{\dag}=(1-\omega^2)U_{\textrm{i}}^{\dag}+(\omega-\omega^2)D_{\textrm{i}}^{\dag}$, and the vacuum state $|0\rangle$ is defined as $d_{\textrm{i}}|0\rangle=0$. The Hamiltonian is diagonalised and  the ground state is unique. This is a trivial case. 
 
\item $\dot{\theta}_1=0$, $\dot{\theta}_2>0.$

In this case the Hamiltonian is:
\begin{eqnarray}
\hat{H}_2&=&-\frac{2\hbar}{3}\dot{\theta}_2\omega^2\sum_{k=1}^{N-1}\left(C_{\textrm{2i}}^{\dagger}C_{\textrm{2i}+1}+C_{\textrm{2i}}C_{\textrm{2i}+1}^{\dagger}\right)\\
&=&\frac{2\hbar}{3}\dot{\theta}_2\sum_{i=1}^{N-1}\left[\tilde{d}_{\textrm{i}}^{\dagger}\tilde{d}_{\textrm{i}}-2\right].\label{YBESU3topo}
\end{eqnarray}
Here the quasiparticle at lattice can be defined as $\tilde{d}_{\textrm{i}}=C_{\textrm{2i}}-\omega^2C_{\textrm{2i}+1}$. The ground states satisfy the condition $\tilde{d}_{\textrm{i}}|\psi\rangle=0$ for $i=1,..., N-1$.  Under the open boundary condition, it shows that the absent operators$C_{\textrm{1}}$, $C_{\textrm{1}}^{\dagger}$,  $C_{\textrm{2N}}$ and $C_{\textrm{2N}}^{\dagger}$ in $\hat{H}_2$ remain unpaired and are the symmetry operators of the Hamiltonian $\hat{H}_2$. Together with the $\omega$-parity operator $P$,  $C_{\textrm{1}}$, $C_{\textrm{1}}^{\dagger}$,  $C_{\textrm{2N}}$ and $C_{\textrm{2N}}^{\dagger}$ lead to the triple degeneracy of ground states which can be categorized according to $P$.  The Hamiltonian has three degenerate ground states: $|\psi_0\rangle$, $|\psi_1\rangle=d^{\dag}|\psi_0\rangle$ and $|\psi_2\rangle=[d^{\dag}]^2|\psi_0\rangle$, where  $d^{\dag}=C_{\textrm{1}}^{\dag}-\omega C_{\textrm{2N}}^{\dag}$. The three ground states $|\psi_0\rangle$, $|\psi_1\rangle$ and $|\psi_2\rangle$ possess the parity $1$, $\omega$ and $\omega^2$, respectively.
\medskip
\end{enumerate}

From the above discussion we see that the $\mathbb{Z}_3$ parafermionic chain is natural generalization of the $\mathbb{Z}_2$ Kitaev model. Next let us  construct the topological invariant for the parafermionic chain and discuss its phase transition. In terms of Fourier transformation, 
\begin{equation}
\begin{split}
&\tilde{U}^{\dag}_{\textrm{k}}={\displaystyle \frac{1}{\sqrt{N}}}{\sum_{\textrm{m}=1}^{\textrm{N}}}e^{{\scriptscriptstyle -\textrm{i}mk}}U_{\textrm{m}}^{\dag}, \, \tilde{U}_{\textrm{k}}={\displaystyle \frac{1}{\sqrt{N}}}\sum_{\textrm{m}=1}^{\textrm{N}}e^{{\scriptscriptstyle \textrm{i}mk}}U_{\textrm{m}};\\
&\tilde{S}^{\dag}_{\textrm{k}}={\displaystyle \frac{1}{\sqrt{N}}}\sum_{\textrm{m}=1}^{\textrm{N}}e^{{\scriptscriptstyle -\textrm{i}mk}}S_{\textrm{m}}^{\dag}, \, \tilde{S}_{\textrm{k}}={\displaystyle \frac{1}{\sqrt{N}}}\sum_{\textrm{m}=1}^{\textrm{N}}e^{{\scriptscriptstyle \textrm{i}mk}}S_{\textrm{m}};\\
&\tilde{D}^{\dag}_{\textrm{k}}={\displaystyle \frac{1}{\sqrt{N}}}\sum_{\textrm{m}=1}^{\textrm{N}}e^{{\scriptscriptstyle -\textrm{i}mk}}D_{\textrm{m}}^{\dag}, \, \tilde{D}_{\textrm{k}}={\displaystyle \frac{1}{\sqrt{N}}}\sum_{\textrm{m}=1}^{\textrm{N}}e^{{\scriptscriptstyle \textrm{i}mk}}D_{\textrm{m}}.
\end{split}
\end{equation}
The parafermionic operators in momentum space can be written in the following form
\begin{eqnarray}
&\tilde{C}_{\textrm{A, k}}^{\dag}={\displaystyle \frac{1}{\sqrt{N}}}{\sum_{\textrm{m}=1}^{\textrm{N}}}e^{{\scriptscriptstyle -\textrm{i}mk}}C_{\textrm{2m-1}}^{\dag}, \, &\tilde{C}_{\textrm{A, k}}={\displaystyle \frac{1}{\sqrt{N}}}\sum_{\textrm{m}=1}^{\textrm{N}}e^{{\scriptscriptstyle \textrm{i}mk}}C_{\textrm{2m-1}};\\
&\tilde{C}_{\textrm{B, k}}^{\dag}={\displaystyle \frac{1}{\sqrt{N}}}{\sum_{\textrm{m}=1}^{\textrm{N}}}e^{{\scriptscriptstyle -\textrm{i}mk}}C_{\textrm{2m}}^{\dag}, \, &\tilde{C}_{\textrm{B, k}}={\displaystyle \frac{1}{\sqrt{N}}}\sum_{\textrm{m}=1}^{\textrm{N}}e^{{\scriptscriptstyle \textrm{i}mk}}C_{\textrm{2m}}.
\end{eqnarray}
Then the equation (\ref{YBEcolorKitaev}) can be expressed in momentum space as
\begin{equation}\label{MHamitonian}
\hat{H}=-\frac{2\hbar}{3}\sum_{\textrm{k}=-\pi}^{\pi}\left[\begin{array}{cc} \tilde{C}^{\dag}_{\textrm{A, k}}, & \tilde{C}^{\dag}_{\textrm{B, k}}\end{array}\right] M \left[\begin{array}{cc} \tilde{C}_{\textrm{A, k}}, & \tilde{C}_{\textrm{B, k}}\end{array}\right]^T,
\end{equation}
where $M$ is $2\times2$ matrix,
\begin{equation}
\begin{split}
&M=\left(\dot{\theta}_1\omega^2+\dot{\theta}_2 e^{-\textrm{i}k}\right)\sigma^{+}+\left(\dot{\theta}_1\omega+\dot{\theta}_2 e^{\textrm{i}k}\right)\sigma^{-};\label{Mmatrix}\\
&\sigma^{+}=\left[\begin{array}{cc}0 & 1\\0 &0\end{array}\right], \quad \sigma^{-}=\left[\begin{array}{cc}0 & 0\\1 &0\end{array}\right].
\end{split}
\end{equation}
Here we note that $[\begin{array}{ccc} \tilde{C}^{\dag}_{\textrm{A, k}}, & \tilde{C}^{\dag}_{\textrm{B, k}}\end{array}]$ is analogous to the Majorana operators in momentum space in 1D Kitaev model. If one define $[\sigma^x, \sigma^y, \sigma^z]$ as a SU(2) space, $M$ can be regarded as a vector in XY-plane of SU(2) space with the basis $\sigma^x$ and $\sigma^y$ ($k\in[-\pi, \pi]$), namely,
\begin{eqnarray}
&&M=\vec{M}\cdot\vec{\sigma},\\
&& \sigma^x=\sigma^++\sigma^-,\, \sigma^y=-\textrm{i}\sigma^++\textrm{i}\sigma^-, \, \sigma^z=\sigma^+\sigma^--\sigma^-\sigma^+,\\
&&\vec{M}=\left(\begin{array}{ccc} -\frac{1}{2}\dot{\theta}_1+\dot{\theta}_2 \cos (k+\frac{2\pi}{3}), & \frac{\sqrt{3}}{2}\dot{\theta}_1-\dot{\theta}_2 \sin (k+\frac{2\pi}{3}), & 0 \end{array}\right),\\
&&\hat{M}=\frac{\vec{M}}{|\vec{M}|}.
\end{eqnarray}
Now the topological invariant for vector $\vec{M}$ can be defined \cite{bernevig2013topological},
\begin{equation}
W=\int^{\pi}_{-\pi}\frac{dk}{4\pi}\epsilon_{\alpha\beta}\hat{M}^{-1}_{\alpha}\frac{\partial\hat{M}_{\beta}}{\partial k}.
\end{equation}

Indeed, the topological invariant $W$ means the winding number of the vector $\vec{M}$ winding around the original point in the first Brillouin zone. In Ref. \cite{fendley2012parafermionic}, the author emphasized that the energy spectrum of parafermionic model can not be obtained simply by Fourier transformation due to the relation in equation (\ref{OmegaCommute}). Here we do not expect to obtain the energy spectrum, but in the momentum basis of $\tilde{C}^{\dag}_{\textrm{A, k}}$ and $\tilde{C}^{\dag}_{\textrm{B, k}}$, the topological winding number of $\mathbb{Z}_3$ parafermionic chain shows the analogous characteristic as the $\mathbb{Z}_2$ Kitaev chain. When $|\dot{\theta}_2|>|\dot{\theta}_1|$, the winding number $W=-1$ corresponds to the topological non-trivial phase. When $|\dot{\theta}_2|<|\dot{\theta}_1|$, the winding number $W=0$ corresponds to the topological trivial phase. In this sense, $|\dot{\theta}_2|=|\dot{\theta}_1|$ is the phase transition point. By calculating the eigenvalues of $M$, we can find that  the ``bulk gap'' closes at $|\dot{\theta}_2|=|\dot{\theta}_1|$ where the ``bulk gap'' closes, the phase transition occurs. Thus we see from the above definition that the critical point of the phase transition $|\dot{\theta}_2|=|\dot{\theta}_1|$ coincides with the $\mathbb{Z}_3$ conformal field theory(CFT)\cite{dotsenko1984critical,zamolodchikov1985nonlocal,mong2014parafermionic}. Obviously, the above properties in our derived $\mathbb{Z}_3$ parafermionic chain are very similar to 1D Kitaev model. However, there are still some differences between  $\mathbb{Z}_2$ and $\mathbb{Z}_3$ models. The critical point of $\mathbb{Z}_2$ Kitaev model can be described by Ising CFT. When Kitaev model is in topological phase, it appears Majorana zero mode with quantum dimension $\sqrt{2}$. While the critical point of $\mathbb{Z}_3$ parafermion model is also described by $\mathbb{Z}_3$ parafermion CFT, but the non-abelian primary fields are not $\mathbb{Z}_3$ parafermion field. There are totally six different quasiparticles in $\mathbb{Z}_3$ parafermion model, three of which possess abelian fields, the vacuum $\mathbb{I}$, parafermion field $\psi$ and $\psi^{\dag}$. Besides, there exist three types of non-abelian fields, $\sigma$, $\sigma^\dag$, $\epsilon$, where $\sigma$ is the spin field and $\epsilon=\sigma\psi$  is the Fibonacci anyon with quantum qimension $\frac{1+\sqrt{5}}{2}$. For example, the $\mathbb{Z}_3$ Read-Rezayi quantum Hall phase supports Fibonacci anyons, which are applicable to universal quantum computation\cite{mong2014parafermionic,read1999beyond,nayak2008non}.

Now let us discuss the generalization of the cyclic chain model from $\mathbb{Z}_3$ to $\mathbb{Z}_N$. To start with, let us introduce the irreducible cyclic representation of SU(N) generators. Under the N-dimensional orthonormal basis $\{|\lambda\rangle\ | \lambda=1,2,...N\}$, akin to SU(3), the ket-bra representation of $N^2-1$ SU(N) generators are
\begin{eqnarray}
&&T_{\textrm{i}}^{(\textrm{m})}=\sum_{a=1}^{N}|a\rangle\langle a+m-1_{\textrm{(mod N)}}|, \quad(i,m=1,2,...N)\\
&&T_{\textrm{i}}^{(\textrm{m})}T_{\textrm{j}}^{(\textrm{n})}=\omega^{(m-1)(j-1)}T^{(m+n-1_{(\textrm{mod N})})}_{i+j+1_{(\textrm{mod N})}},\label{SUNrelation}
\end{eqnarray}
with $\omega$ root of unity $\omega^N=1$ and $T_{1}^{(1)}$ identity operator. Then the representation of Temperley-Lieb algebra $T_{i}$ on (i, i+1)-th sites under the cyclic $SU(N)$ operators is expressed as
\begin{equation}\label{TN}
T_i=\frac{1}{\sqrt{N}}\sum_{k=1}^{N}\left[T_{\textrm{x}}^{(\textrm{m})}\otimes T_{\textrm{y}}^{(\textrm{m})}\right]_{i,i+1}^k,
\end{equation}
with $x,y,m,n$  arbitrary certain integer given from 2 to N. Based on the algebraic relation in equation (\ref{SUNrelation}), it is easy to check that $T_i$ in equation (\ref{TN}) satisfies TLA with quantum dimension $d=\sqrt{N}$,
\begin{equation}\label{TLAN}
\begin{split}
&T_{\textrm{i}}^2=dT_{\textrm{i}},\quad d=\sqrt{N},\\
&T_{\textrm{i}}T_{\textrm{i}\pm1}T_{\textrm{i}}=T_{\textrm{i}},\\
&T_{\textrm{i}}T_{\textrm{j}}=T_{\textrm{j}}T_{\textrm{i}}, \quad |i-j|>1.
\end{split}
\end{equation}

From the view point of rational Yang-Baxterization of Temperley-Lieb algebra,  when $d\leq2$, i.e. $N\leq4$, the parameter in the solution of YBE is real  and the corresponding $\breve{R}$-matrix is unitary and can be viewed as unitary evolution operator of a quantum system. That is, we can obtain $\mathbb{Z}_4$ parafermion chain from YBE in an similar way as $\mathbb{Z}_3$.  While $d>2$, the parameter in the solution of YBE is imaginary(see Supplementary), hence the $\breve{R}$ is not unitary and cannot be viewed as an ideal evolution operator. One can still construct $\mathbb{Z}_{N>4}$ parafermion chain, but the chain does not come from the rational Yang-Baxterization.

\bigskip

\noindent \textbf{Fermionic representation of $\mathbb{Z}_3$ parafermionic model.}
In this section we express the $\mathbb{Z}_3$ parafermionic chain in terms of fermions. For the $3\times3$ matrices in equation (\ref{su3ladder}), we choose three orthonormal basis, $|r\rangle=r^{\dag}|\textrm{vac}\rangle$, $|g\rangle=g^{\dag}|\textrm{vac}\rangle$ and $|b\rangle=b^{\dag}|\textrm{vac}\rangle$. Here $|\textrm{vac}\rangle$ represents the vacuum state,  $r^{\dag}$, $g^{\dag}$ and $b^{\dag}$ are three types of fermions and satisfy the fermionic commutation conditions
\begin{equation}
\{x^{\dag}, y\}=\delta_{xy}, \quad \{x^{\dag}, y^{\dag}\}=\{x, y\}=0, \qquad (x, y= r, g, b),
\end{equation}
with the constraint of the occupation number of the fermions on each site
\begin{equation}\label{FerCon}
r^{\dag}r+g^{\dag}g+b^{\dag}b=1.
\end{equation} 
Then equation (\ref{FerCon}) means the only one occupied fermion on the site for either $r^{\dag}$ or $g^{\dag}$ or $b^{\dag}$. Considering equation (\ref{su3ladder}) the SU(3) ladder operators can be written as
\begin{equation}\label{fermionusd}
\begin{split}
&u^{+}=r^{\dag}g, \, s^{+}=\omega g^{\dag}b, \, d^{+}=\omega^2 b^{\dag}r,\\
&u^-=g^{\dag}r, \, s^-=\omega^2 b^{\dag}g, \, d^-=\omega r^{\dag}b.
\end{split}
\end{equation}
In the basis of $|r\rangle$, $|g\rangle$ and $|b\rangle$, we have the relation
\begin{equation}
w^{\dag}xy^{\dag}z|n\rangle=\delta_{xy}w^{\dag}z|n\rangle-w^{\dag}z^{\dag}xy|n\rangle=\delta_{xy}w^{\dag}z|n\rangle. \quad(w, x, y, z,n\in\{r, g, b\})
\end{equation}
It can be proved that the operators in equation (\ref{fermionusd}) satisfy the same relations as for those in equation (\ref{usdrelation}) (see Supplementary). In other words, the fermionic representation of the SU(3) operators also satisfy the matrix multiplication.  Similarly, the nonlocal operators $U^{\dag}$, $S^{\dag}$ and $D^{\dag}$ can be expressed as

\begin{equation}\label{USDfermion}
\begin{split}
&U_{\textrm{n}}^{\dagger}=\left[\prod_{\textrm{i}=1}^{\textrm{n}-1}[r^{\dag}_\textrm{i}r_\textrm{i}+\omega g^{\dag}_\textrm{i}g_\textrm{i} +\omega^2 b^{\dag}_\textrm{i}b_\textrm{i} ]\right]r^{\dag}_{\textrm{n}}g_{\textrm{n}},\quad U_{\textrm{n}}=\left[\prod_{\textrm{i}=1}^{\textrm{n}-1}[r^{\dag}_\textrm{i}r_\textrm{i}+\omega^2 g^{\dag}_\textrm{i}g_\textrm{i} +\omega b^{\dag}_\textrm{i}b_\textrm{i} ]\right]g^{\dag}_{\textrm{n}}r_{\textrm{n}};\\
&S_{\textrm{n}}^{\dagger}=\left[\prod_{\textrm{i}=1}^{\textrm{n}-1}[r^{\dag}_\textrm{i}r_\textrm{i}+\omega g^{\dag}_\textrm{i}g_\textrm{i} +\omega^2 b^{\dag}_\textrm{i}b_\textrm{i} ]\right]\omega g^{\dag}_{\textrm{n}}b_{\textrm{n}},\quad S_{\textrm{n}}=\left[\prod_{\textrm{i}=1}^{\textrm{n}-1}[r^{\dag}_\textrm{i}r_\textrm{i}+\omega^2 g^{\dag}_\textrm{i}g_\textrm{i} +\omega b^{\dag}_\textrm{i}b_\textrm{i} ]\right]\omega^2 b^{\dag}_{\textrm{n}}g_{\textrm{n}};\\
&D_{\textrm{n}}^{\dagger}=\left[\prod_{\textrm{i}=1}^{\textrm{n}-1}[r^{\dag}_\textrm{i}r_\textrm{i}+\omega g^{\dag}_\textrm{i}g_\textrm{i} +\omega^2 b^{\dag}_\textrm{i}b_\textrm{i} ]\right]\omega^2 b^{\dag}_{\textrm{n}}r_{\textrm{n}},\quad D_{\textrm{n}}=\left[\prod_{\textrm{i}=1}^{\textrm{n}-1}[r^{\dag}_\textrm{i}r_\textrm{i}+\omega^2 g^{\dag}_\textrm{i}g_\textrm{i} +\omega b^{\dag}_\textrm{i}b_\textrm{i} ]\right]\omega r^{\dag}_{\textrm{n}}b_{\textrm{n}}.
\end{split}
\end{equation}
Here we do not need to require whether the commutation relation between two operators on different sites is fermionic or bosonic, since the non local operators $U$, $S$ and $D$  are always even power of the fermionic operators. 

Making use of the fermionic representation of $U^{\dag}$, $S^{\dag}$ and $D^{\dag}$  we find that $T_{\textrm{i}}'$ given by equation (\ref{Crepresentation}) also satisfies T-L algebraic relation.  Then the $\mathbb{Z}_3$ parafermionic chain is rewritten as
\begin{eqnarray}\label{rgbparafermion}
\hat{H}&=&-\frac{2\hbar}{3}\omega^2\sum_{i=1}^{N}{\Big [ }\dot{\theta}_1\left(C_{\textrm{2i}-1}^{\dagger}C_{\textrm{2i}}+C_{\textrm{2i}-1}C_{\textrm{2i}}^{\dagger}\right)+\dot{\theta}_2\left(C_{\textrm{2i}}^{\dagger}C_{\textrm{2i}+1}+C_{\textrm{2i}}C_{\textrm{2i}+1}^{\dagger}\right)\Big]\nonumber\\
&=&-\frac{2\hbar}{3}\sum_{i=1}^{N}{\Big [ }\dot{\theta}_1\left(3g^{\dag}_\textrm{i}g_\textrm{i}-1\right)+\dot{\theta}_2\left(-\tilde{d}_{\textrm{i}}^{\dag}\tilde{d}_{\textrm{i}}+2\right)\Big],
\end{eqnarray}
where $\tilde{d}_{\textrm{i}}^{\dag}=r^{\dag}_\textrm{i}g_\textrm{i}+ g^{\dag}_\textrm{i}b_\textrm{i} + b^{\dag}_\textrm{i}r_\textrm{i}-\omega(r^{\dag}_\textrm{i+1}g_\textrm{i+1}+ g^{\dag}_\textrm{i+1}b_\textrm{i+1} + b^{\dag}_\textrm{i+1}r_\textrm{i+1})$.  For the topological non-trivial case $\dot{\theta}_1=0$, $\dot{\theta}_2>0$, $\tilde{d}_{\textrm{i}}^{\dag}$ in equation (\ref{rgbparafermion}) shows the rotation symmetry of $r^{\dag}$, $g^{\dag}$ and $b^{\dag}$ for each sited SU(3) spin. For the topological trivial case $\dot{\theta}_1>0$, $\dot{\theta}_2=0$, equation (\ref{rgbparafermion}) shows that the ground state corresponds to the full occupation of fermion $g^{\dag}_{\textrm{i}}$ on each site. Unlike the topological non-trivial case, there is no cyclic permutation symmetry of $r^{\dag}$, $g^{\dag}$ and $b^{\dag}$ on each SU(3) spin site.  There are totally three types of parafermions on the i-th SU(3) spin site, $F_i^{\dag}$, $G_{i}^{\dag}$ and $\omega F_iG_i$ (see equation (\ref{fgd})), but we only choose two of the three types of parafermions to represent Temperley-Lieb algebra in equation (\ref{FGTLA}). Hence there are three different ways to choose the representation of TLA as well as the $\mathbb{Z}_3$ parafermionic model. Each way corresponds to one type of ground state occupation($r^{\dag}$ or $g^{\dag}$ or $b^{\dag}$).   
\bigskip

\noindent \textbf{Algebra of triple degenerate model.}
In previous sections, we have discussed the $\mathbb{Z}_3$ parafermionic model  and the physical consequences. There appears the triple degeneracy in ground state and the emergence of tripling corresponds to the topological phase of the Hamiltonian. It can be regarded as the extension of the algebra of Majorana doubling pointed out in Ref. \cite{lee2013algebra}. In this section, we shall show the algebra of triple degeneracy at each energy level due to the 3-cyclic and give its intuitive explanation.

Firstly let us construct 3-body Hamiltonian based on the 3-body S-matrix constrained by YBE. It is well known that the physical meaning of $\breve{R}_{\textrm{i}}(\theta)$ is 2-body S-matrix. YBE means that a 3-body S-matrix can be decomposed into three 2-body S-matrices in the following way
\begin{eqnarray}
 \breve{R}_{123}(\theta_1,\theta_2,\theta_3)
&=&\breve{R}_{12}(\theta_1) \breve{R}_{23}(\theta_2) \breve{R}_{12}(\theta_3)\nonumber\\
&=&\breve{R}_{23}(\theta_3) \breve{R}_{12}(\theta_2) \breve{R}_{23}(\theta_1).
 \end{eqnarray}
Here we note that due to the constraint of equation (\ref{YBEangularrelation}),  only two of the three parameters $\theta_1$, $\theta_2$ and $\theta_3$ are free. Suppose $\theta_1$ and $\theta_2$ are time dependent, then the 3-body Hamiltonian can be obtained from equation (\ref{SchrodingerEquation}) (see Supplementary)
 \begin{eqnarray}
 \hat{H}_{123}=\omega^2[\alpha(C_{\textrm{1}}^{\dagger}C_{\textrm{2}}+C_{\textrm{1}}C_{\textrm{2}}^{\dagger})+\beta(C_{\textrm{1}}^{\dagger}C_{\textrm{3}}+C_{\textrm{1}}C_{\textrm{3}}^{\dagger})+\gamma(C_{\textrm{2}}^{\dagger}C_{\textrm{3}}+C_{\textrm{2}}C_{\textrm{3}}^{\dagger})]+\kappa(C_{\textrm{1}}^{\dagger}C_{\textrm{2}}^{\dagger}C_{\textrm{3}}^{\dagger}+C_{\textrm{1}}C_{\textrm{2}}C_{\textrm{3}}),
 \end{eqnarray}
 where $\alpha$, $\beta$, $\gamma$ and $\kappa$ are real parameters depending on $\theta_1$ and $\theta_2$. By making inverse Jordan-Wigner transformation for SU(3) to transform $C_{\textrm{i}}$'s back into SU(3) spin sites, one can show that there are only two independent symmetry operators (see Supplementary)
 \begin{eqnarray}
 &&P=C_1C_2^{\dag}C_3C_4^{\dag},\\
 &&\Gamma=\omega C_1^{\dag}C_2C_3^{\dag}.
 \end{eqnarray}
 Then the complete set of the algebra for the Hamiltonian is 
 \begin{eqnarray}
 &&[\hat{H}_{123}, P]=0, \, [\hat{H}_{123}, \Gamma]=0;\\
 &&\Gamma P=\omega P\Gamma, \, P^3=1, \, \Gamma^3=1.\label{GPcommute}
 \end{eqnarray}
Here $P$ represents the $\omega$-parity operator. Now we turn to the analysis of the degeneracy of the Hamiltonian.  From equation (\ref{GPcommute}), $\Gamma$ transforms the common eigenstates $|\psi\rangle$ of $\hat{H}_{123}$ and $P$ to the following form:
\begin{eqnarray}
&&P|\psi\rangle=a|\psi\rangle, \quad (a=1,\omega, \omega^2);\\
&&P[\Gamma|\psi\rangle]=\omega^2\Gamma P|\psi\rangle=a\omega^2\Gamma|\psi\rangle;\\
&&P[\Gamma^2|\psi\rangle]=\omega\Gamma^2 P|\psi\rangle=a\omega\Gamma^2|\psi\rangle.
\end{eqnarray} 
Because $\Gamma$ commutes with the Hamiltonian $\hat{H}_{123}$, the above three states have the same energy with different $\omega$-parity. As a consequence the Hamiltonian possesses triple degeneracy on all energy levels. In this sense, we conclude that $\mathbb{Z}_2$ parity leads to Majorana doubling, whereas the $\mathbb{Z}_3$ $\omega$-parity leads to the tripling.

\section*{DISCUSSION}

Before ending the paper, we would like to make some comments and discussions. 

%
1) In our SU(3) models, only the three operators $U^{\dag}$, $S^{\dag}$ and $D^{\dag}$ are basic operators that are extention of the spinless fermion $a^{\dag}$ and $a$ in SU(2) Majorana models. $U$, $S$ and $D$ can be expressed by $U^{\dag}$, $S^{\dag}$ and $D^{\dag}$(see equation (\ref{USDrelations})).

2) Our results can be regarded as the direct generalization of $\mathbb{Z}_2$ Majorana models. The comparison between $\mathbb{Z}_2$ Majorana fermion and $\mathbb{Z}_3$  parafermion is shown in the Table \ref{tab:Compare23}. In $\mathbb{Z}_2$ case, the eigenvalues of the symmetry operators are $\pm1$ whereas the eigenvalues turn into $1$, $\omega$ and $\omega^2$ for $\mathbb{Z}_{3}$ symmetry operators.  We see from the Table \ref{tab:Compare23} that the exchange of $\mathbb{Z}_{3}$ quasiparticles $C_{\textrm{i}}$ emerges an extra $\omega$ or $\omega^2$ phase factor instead of $-1$.   

3) The topological case of $\mathbb{Z}_3$ parafermionic model has been obtained, which corresponds to the triple degenerate ground states and the non-trivial topological winding number. Although the $\mathbb{Z}_3$ parafermionic model consists of SU(3) spin, the matrix $M$ of the Hamiltonian in momentum space still forms SU(2)(see equation (\ref{Mmatrix})). Hence we can follow the original Kitaev model to define the topological winding number for the $\mathbb{Z}_3$ parafermionic model. The $\mathbb{Z}_3$ parafermionic model is formed by three types of fermions $r^\dag$, $g^\dag$ and $b^\dag$. For topological trivial case, the ground state corresponds to the full occupation of $g$-fermion. For topological non-trivial case, the ground state shows the cyclic rotational symmetry of  $r$, $g$ and $b$-fermions. This may be helpful to realize the $\mathbb{Z}_3$ parafermionic model experimentally.

4) To give an intuitive explanation about the triple degeneracy, we construct the 3-body Hamiltonian $\hat{H}_{123}$ based on YBE. This model is the generalization of the 3-MF model pointed by Lee and Wilczek. Two independent symmetry operators $\Gamma$ and $P$ of the Hamiltonian have been found and we show that all the energy level of $\hat{H}_{123}$ possesses triple degeneracy. In the process of constructing triple degeneracy, the 3-cyclic property of the $\mathbb{Z}_3$ symmetry operators plays the crucial role. 

5) Both of the two derived models are based on the $9\times9$ braid matrix $B_{\textrm{i}}$ as well as the solution $\breve{R}_{\textrm{i}}(\theta)$ of YBE. Regarding $\breve{R}_{\textrm{i}}(\theta)$ as the time dependent unitary evolution of 2-body interaction, one can construct the local 2-body interacting Hamiltonian. The 2-body Hamiltonian is then extended to the desired $\mathbb{Z}_3$ parafermionic chain. To obtain the 3-body Hamiltonian, we suppose that 3-body S-matrix can be decomposed into three 2-body S-matrices via YBE that is acceptable in low-energy physics. The advantage is in that the 3-body Hamiltonian inherits the $\omega$-parity symmetry from 2-body Hamiltonian. In other words, the $\omega$-parity symmetry of $\breve{R}(\theta)$ preserves the symmetry properties of $\mathbb{Z}_3$ parafermionic model and 3-body Hamiltonian. This is the important role of YBE plays in obtaining the desired models.

To summarize, we extend the $\mathbb{Z}_2$ Kitaev model to $\mathbb{Z}_3$ parafermionic model. Due to the 3-fold 3-cyclic, $\mathbb{Z}_3$ model possesses triple degeneracy. But both $\mathbb{Z}_2$ and $\mathbb{Z}_3$ models  have the similar topological phase transition scheme obtained from YBE. In this sense, YBE is a powerful tool for generating new models.  How to make full use of YBE to generate more meaningful models is still a challenge problem.

\bibliographystyle{unsrt}

\begin{thebibliography}{10}

\bibitem{kitaev2001unpaired}
Kitaev, A. Y. 
\newblock {  Unpaired Majorana fermions in quantum wires},
\newblock \href{http://iopscience.iop.org/1063-7869/44/10S/S29}{{\em Phys. Usp.} {\bf 44,} 131 (2001).}

\bibitem{ivanov2001non}
Ivanov, D. A. 
\newblock {  Non-Abelian statistics of half-quantum vortices in $\mathit{p}$-wave superconductors},
\newblock \href{http://journals.aps.org/prl/abstract/10.1103/PhysRevLett.86.268}{{\em Phys. Rev. Lett.} {\bf 86,} 268 (2001).}

\bibitem{kitaev2006anyons}
Kitaev, A. Y. 
\newblock {  Anyons in an exactly solved model and beyond},
\newblock  \href{http://www.sciencedirect.com/science/article/pii/S0003491605002381}{{\em Ann. Phys.} {\bf 321,} 2 (2006).}

\bibitem{wilczek2009majorana}
Wilczek, F. 
\newblock {  Majorana returns},
\newblock \href{http://www.nature.com/nphys/journal/v5/n9/abs/nphys1380.html}{{\em Nature Phys.} {\bf 5,} 614 (2009).}

\bibitem{leijnse2012introduction}
Leijnse, M. \& Flensberg, K. 
\newblock{  Introduction to topological superconductivity and Majorana fermions},
\newblock \href{http://iopscience.iop.org/0268-1242/27/12/124003}{{\em Semicond. Sci. Tech.} {\bf 27,} 124003 (2012).}

\bibitem{alicea2011non}
Alicea, J. {\em et al.}
\newblock {  Non-Abelian statistics and topological quantum information processing in {1D} wire networks},
\newblock \href{http://www.nature.com/nphys/journal/v7/n5/abs/nphys1915.html}{{\em Nat. Phys.} {\bf 7,} 412 (2011).}

\bibitem{lee2013algebra}
Lee, J.  \& Wilczek, F. 
\newblock {  Algebra of Majorana doubling},
\newblock \href{http://journals.aps.org/prl/abstract/10.1103/PhysRevLett.111.226402}{{\em Phys. Rev. Lett.} {\bf 111,} 226402 (2013).}

\bibitem{yu2015more}
Yu, L. W.  \& Ge, M. L. 
\newblock{  More about the doubling degeneracy operators associated with Majorana fermions and Yang-Baxter equation},
\newblock \href{http://www.nature.com/srep/2015/150129/srep08102/full/srep08102.html?WT.ec_id=SREP-20150203}{{\em Sci. Rep.} {\bf 5,} 8102 (2015).}

\bibitem{yang1967some}
Yang, C. N. 
\newblock {  Some exact results for the many-body problem in one dimension with repulsive delta-function interaction},
\newblock \href{http://journals.aps.org/prl/abstract/10.1103/PhysRevLett.19.1312}{{\em Phys. Rev. Lett.} {\bf 19,} 1312 (1967).}

\bibitem{baxter1972partition}
Baxter, R. J. 
\newblock {  Partition function of the eight-vertex lattice model},
\newblock \href{http://www.sciencedirect.com/science/article/pii/0003491672903351}{{\em Ann. Phys.} {\bf 70,} 193 (1972).}

\bibitem{faddeev1982integrable}
Faddeev, L. D. 
\newblock {  Integrable models in (1+1)-dimensional quantum field theory}, 
\newblock \href{}{{\em Les Houches Lectures} {\bf 39,} 561 (1982).}

\bibitem{kulish1981lecture}
Kulish, P.   \&  Sklyanin, E. 
\newblock {  Lecture Notes in Physics, Vol. 151 Springer, Berlin (1982)},
\newblock in {\em Yang-Baxter equation in integrable systems} 61 {(World Scientific, Singapore,1990)}.

\bibitem{korepin1997quantum}
Korepin, V. E. 
\newblock { \em Quantum inverse scattering method and correlation functions}
\newblock {(Cambridge University Press, 1997).}


\bibitem{takhtajan1989lectures}
Takhtajan, L. A. 
\newblock {  Lectures on Quantum Groups},
\newblock in {\em Lectures on Math. Phys.}, edited by Ge, M. L.  \&  Zhao, B. H.
  (World Scientific, Singapore, 1989).

\bibitem{chen2007braiding}
Chen, J. L., Xue, K.  \&  Ge, M. L. 
\newblock {  Braiding transformation, entanglement swapping, and Berry phase in
  entanglement space},
\newblock \href{http://journals.aps.org/pra/abstract/10.1103/PhysRevA.76.042324}{{\em Phys. Rev. A} {\bf 76,} 042324 (2007).}

\bibitem{chen2008berry}
Chen, J. L.,  Xue, K.  \&  Ge, M. L.
\newblock {  Berry phase and quantum criticality in Yang-Baxter systems},
\newblock \href{http://www.sciencedirect.com/science/article/pii/S0003491608000973}{{\em Ann. Phys.} {\bf 323,} 2614 (2008).}

\bibitem{ge2012yang}
Ge, M. L.  \& Xue, K. 
\newblock {  Yang-Baxter equations in quantum information},
\newblock \href{http://www.worldscientific.com/doi/pdf/10.1142/S0217979212430072}{{\em  Int. J. Mod. Phys. B} {\bf 26,} 27n28 (2012).}

\bibitem{yu2014factorized}
 Yu, L. W., Zhao, Q.  \&  Ge, M. L. 
\newblock {  Factorized three-body S-matrix restrained by Yang-Baxter equation
  and quantum entanglements},
\newblock \href{http://www.sciencedirect.com/science/article/pii/S0003491614001286}{{\em  Ann. Phys.} {\bf 348,} 106 (2014).}

\bibitem{kauffman2004braiding}
Kauffman, L. H.   \&  Lomonaco Jr, S. J.  
\newblock {  Braiding operators are universal quantum gates},
\newblock \href{http://iopscience.iop.org/1367-2630/6/1/134}{{\em New J. Phys.} {\bf 6,} 134 (2004).}

\bibitem{Rowell2009localization}
Rowell, E. C.  \&  Wang, Z. H. 
\newblock {  Localization of unitary braid group representations},
\newblock \href{http://link.springer.com/article/10.1007/s00220-011-1386-7}{{\em Commun. Math. Phys.} {\bf 311,} 595 (2012).}

\bibitem{Temperley1971relations}
Temperley, H. V \& Lieb, E. L. 
\newblock { Relations between the `percolation' and `coloring' problem and other graph-theoretical problems associated with regular planar lattices: some exact results for the `percolation' problem},
\newblock \href{http://rspa.royalsocietypublishing.org/content/322/1549/251.short}{{\em Proc. R. Soc.  Lon. A} {\bf 322,} 1549 (1971).}


\bibitem{jimbobook}
Jimbo, M.
\newblock{ {\em Yang-Baxter equation in Integral Systems} (World Scientific, Singapore, 1990).}

\bibitem{Kac1994infinite}
Kac, V. G.
\newblock {{\em Infinite-dimensional Lie algebras} (Cambridge University Press, 1994).}

\bibitem{fendley2012parafermionic}
Fendley, P.
\newblock{Parafermionic edge zero modes in $Z_n$ -invariant spin chains},
\newblock  \href{http://stacks.iop.org/1742-5468/2012/i=11/a=P11020}{{\em J. Stat. Mech: Theory and Experiment} {\bf 11,} 11020 (2012).}

\bibitem{liu2013generalized}
 Liu, M., Bai, C. M., Ge, M. L.  \&  Jing, N. H. 
\newblock {  Generalized Bell states and principal realization of the Yangian} $Y(sl_n)$,
\newblock \href{http://scitation.aip.org/content/aip/journal/jmp/54/2/10.1063/1.4789317}{{\em J. Math. Phys.} {\bf 54,} 021701 (2013).}

\bibitem{Greite2007exact}
Greiter, M., Rachel, S.  \&  Schuricht, D. 
\newblock {  Exact results for su(3) spin chains: Trimer states, valence bond solids, and their parent Hamiltonians},
\newblock \href{http://journals.aps.org/prb/abstract/10.1103/PhysRevB.75.060401}{{\em Phys. Rev. B} {\bf 75,} 060401 (2007).}

\bibitem{cobanera2014fock}
Cobanera, E.  \&  Ortiz, G. 
\newblock {  Fock parafermions and self-dual representations of the braid group},
\newblock \href{http://journals.aps.org/pra/abstract/10.1103/PhysRevA.89.012328}{{\em Phys. Rev. A} {\bf 89,} 012328 (2014).}

\bibitem{wiegmann1994bethe}
Wiegmann, P. B.   \&  Zabrodin, A. V. 
\newblock {  Bethe-ansatz for the Bloch electron in magnetic field},
\newblock  \href{http://journals.aps.org/prl/abstract/10.1103/PhysRevLett.72.1890}{{\em Phys. Rev. Lett.} {\bf 72,} 1890 (1994).}


\bibitem{stephen1972pseudo}
Stephen, M. J. \& Mittag, L.
\newblock{Pseudo-Hamiltonians for the Potts model at the critical point},
\newblock \href{http://www.sciencedirect.com/science/article/pii/0375960172909279}{{\em Phys. Lett. A} {\bf 41,}  357 (1972).}

\bibitem{fateev1982self}
Fateev, V. A. \& Zamolodchikov, A. B.
\newblock{Self-dual solutions of the star-triangle relations in $Z_n$-models},
\newblock \href{http://www.sciencedirect.com/science/article/pii/0375960182907368}{{\em Phys. Lett. A} {\bf 92,}  37 (1982).}

\bibitem{baxter1988new}
Baxter, R. J., Perk, J. H. \& Au-Yang, H.
\newblock{New solutions of the star-triangle relations for the chiral Potts model},
\newblock \href{http://www.sciencedirect.com/science/article/pii/0375960188908961}{{\em Phys. Lett. A} {\bf 128,}  138 (1988).}

  
\bibitem{bernevig2013topological}
Bernevig, B. A.  \&  Hughes, T. L. 
\newblock{\em Topological insulators and topological superconductors} 205 (Princeton University Press, 2013).

\bibitem{dotsenko1984critical}
Dotsenko, Vl S. 
\newblock{Critical behaviour and associated conformal algebra of the $Z_3$ Potts model},
\newblock \href{http://www.sciencedirect.com/science/article/pii/0550321384901482} {{\em Nucl. Phys. B} {\bf 235,} 54 (1984).}

\bibitem{zamolodchikov1985nonlocal}
Zamolodchikov, A. B. \& Fateev, V. A. 
\newblock{Nonlocal (parafermion) currents in two-dimensional conformal quantum field theory and self-dual critical points in $Z_N$-symmetric statistical systems},
\newblock \href{http://jetp.ac.ru/cgi-bin/dn/e_062_02_0215.pdf} {{\em Sov. Phys. JETP} {\bf 62,} 215 (1985).}

\bibitem{mong2014parafermionic}
Mong, R. S. {\em et al.}
\newblock{\em Parafermionic conformal field theory on the lattice},
\newblock \href{http://iopscience.iop.org/article/10.1088/1751-8113/47/45/452001/meta;jsessionid=7DA187D3DA2398D0C30529057A97330E.c1.iopscience.cld.iop.org}{{\em J. Phys. A} {\bf 47,} 452001 (2014).}

\bibitem{read1999beyond}
Read, N. \& Rezayi, E.
\newblock{\em Beyond paired quantum Hall states: parafermions and incompressible states in the first excited Landau level},
\newblock \href{http://journals.aps.org/prb/abstract/10.1103/PhysRevB.59.8084}{{\em Phys. Rev. B} {\bf 59,} 8084 (1999).}

\bibitem{nayak2008non}
Nayak, C. {\em et al.}
\newblock{\em Non-Abelian anyons and topological quantum computation},
\newblock \href{http://journals.aps.org/rmp/abstract/10.1103/RevModPhys.80.1083}{{\em Rev. Mod. Phys.} {\bf 80,} 1083 (2008).}

\end{thebibliography}

\vspace{3mm}

\indent{\bf Acknowledgments}
We thank Prof. Cheng-Ming Bai who provides with the Kac's principal representation of SU(3) to the authors. We also thank Dr. Ruo-Yang Zhang for helpful discussion. We are grateful to Prof. F. Wilczek who suggests the authors to find new type of Hamiltonian with $\Gamma$ symmetry. This work is in part supported by NSF of China (Grant No. 11475088).

\vspace{3mm}

{\bf Author contributions}
M.L.G. and L.W.Y. proposed the idea, L.W.Y. performed the calculation and derivation, L.W.Y. and M.L.G. prepared the manuscript, both of the authors contributed equally to this paper and reviewed the manuscript.

\vspace{3mm}

{\bf Additional information}

\textbf{Competing financial interests:} The authors declare no
competing financial interests.

\end{document}


\title{Supplementary for `` $\mathbb{Z}_3$ parafermionic chain emerging from Yang-Baxter equation''}

\author{Li-Wei Yu}
\affiliation{Theoretical Physics Division, Chern Institute of Mathematics, Nankai University,
 Tianjin 300071, China}


\author{Mo-Lin Ge}
\affiliation{Theoretical Physics Division, Chern Institute of Mathematics, Nankai University, Tianjin 300071, China}

\maketitle

\section{$\omega$-cyclic SU(3) generator in Yang-Baxter equation}
The $\omega$-cyclic representation of SU(3) generators are (except $T_{1}^{(1)}$)[1]:
\begin{eqnarray*}
&&T_{1}^{(1)}=\left[\begin{array}{ccc} 1 & 0 & 0 \\ 0 & 1 & 0 \\ 0 & 0 & 1\end{array}\right], T_{1}^{(2)}=\left[\begin{array}{ccc} 0 & 1 & 0 \\ 0 & 0 & 1 \\ 1 & 0 & 0\end{array}\right], T_{1}^{(3)}=\left[\begin{array}{ccc} 0 & 0 & 1 \\ 1 & 0 & 0 \\ 0 & 1 & 0\end{array}\right];\\
&&T_{2}^{(1)}=\left[\begin{array}{ccc} 1 & 0 & 0 \\ 0 & \omega & 0 \\ 0 & 0 & \omega^2 \end{array}\right], T_{2}^{(2)}=\left[\begin{array}{ccc} 0 & 1 & 0 \\ 0 & 0 & \omega \\ \omega^2 & 0 & 0\end{array}\right], T_{2}^{(3)}=\left[\begin{array}{ccc} 0 & 0 & 1 \\ \omega & 0 & 0 \\ 0 & \omega^2 & 0\end{array}\right];\\
&&T_{3}^{(1)}=\left[\begin{array}{ccc} 1 & 0 & 0 \\ 0 & \omega^2 & 0 \\ 0 & 0 & \omega \end{array}\right], T_{3}^{(2)}=\left[\begin{array}{ccc} 0 & 1 & 0 \\ 0 & 0 & \omega^2 \\ \omega & 0 & 0\end{array}\right], T_{3}^{(3)}=\left[\begin{array}{ccc} 0 & 0 & 1 \\ \omega^2 & 0 & 0 \\ 0 & \omega & 0\end{array}\right].
\end{eqnarray*}
and they satisfy the following algebraic relations
\begin{eqnarray}
&&T_{\textrm{i}}^{(\textrm{m})}T_{\textrm{j}}^{(\textrm{n})}=\omega^{\textrm{(m-1)(j-1)}}T_{a}^{(b)} \quad (\omega=e^{\textrm{i}\frac{2\pi}{3}}),\label{SU3relation}\\
&&a=\textrm{i+j-1}(\textbf{mod} \, 3), \, b=\textrm{m+n-1} (\textbf{mod} \, 3). \nonumber
\end{eqnarray}

The braid operator can be parametrized to yield the solution of YBE by means of Yang-Baxterization[2]. Now let us recall the standard method of Yang-Baxterization. The YBE reads,
\begin{equation}
\breve{R}_{\textrm{i}}(\mu) \breve{R}_{\textrm{i}+1}(f(\mu, \nu)) \breve{R}_{\textrm{i}}(\nu)=\breve{R}_{\textrm{i}+1}(\nu) \breve{R}_{\textrm{i}}(f(\mu, \nu)) \breve{R}_{\textrm{i}+1}(\mu),
\end{equation}
and the solution is
\begin{eqnarray}
&&\breve{R}_{\textrm{i}}(\mu)=\rho(\mu)[1+G(\mu)T_{\textrm{i}}],\smallskip\\
&&G(\mu)=\frac{\mu}{a_0-d\mu/2},\\
&&f(\mu,\nu)=\frac{\mu+\nu}{1+\beta^2\mu\nu}, \, \beta^2=\frac{d^2-4}{(2a_0)^2}.
\end{eqnarray}
Here  $T_{\textrm{i}}$ is Temperley-Lieb algebra(TLA) generator, $d$ represents the loop value of TLA and $a_0$ is a free parameter. If we express $T_{\textrm{i}}$ in terms of the known braid operator $B_{\textrm{i}}$, then the solution of YBE can be obtained. In this paper, the T-L  generator associated with braid operators $B_{\textrm{i}}$ can be expressed as follows,
\begin{eqnarray}\label{SU3TgeneartorAppendix}
T_{\textrm{i}}&=&e^{\textrm{i}\frac{\pi}{6}}(I^{\otimes2}+B_{\textrm{i}})\\
&=&\frac{1}{\sqrt{3}}(I^{\otimes2}+\omega T_{3}^{(2)}\otimes T_{3}^{(3)}+\omega^2 T_{2}^{(3)}\otimes T_{2}^{(2)})_{\textrm{i,i+1}}.\nonumber
\end{eqnarray}
$T_{\textrm{i}}$ satisfy T-L algebra with the loop value $d=\sqrt{3}$,
\begin{equation}
\begin{split}
&T_{\textrm{i}}^2=dT_{\textrm{i}}, d=\sqrt{3}\\
&T_{\textrm{i}}T_{\textrm{i}\pm1}T_{\textrm{i}}=T_{\textrm{i}},\\
&T_{\textrm{i}}T_{\textrm{j}}=T_{\textrm{j}}T_{\textrm{i}}, \quad |i-j|>1.
\end{split}
\end{equation}
Based on the algebraic relation in equation (1), we can verify these relations by direct calculation.

After the replacement
\begin{eqnarray}
&&\tan\theta_1=\beta\frac{\textrm{i}d}{\sqrt{d^2-4}}u,\\
&&\tan\theta_2=\beta\frac{\textrm{i}d}{\sqrt{d^2-4}}f(u,v),\\
&&\tan\theta_3=\beta\frac{\textrm{i}d}{\sqrt{d^2-4}}v,  \,(d=\sqrt{3})
\end{eqnarray}
the solution $\breve{R}_{\textrm{i}}(u)$ is rewritten as
\begin{equation}
\breve{R}_{\textrm{i}}(\theta)=\cos\theta  I+\textrm{i}\sin\theta(\frac{2}{\sqrt{3}}T_{\textrm{i}}-I),
\end{equation}
with the angular relation
\begin{equation}\label{YBEangularrelationAppendix}
\tan\theta_2=\frac{\tan\theta_1+\tan\theta_3}{1+\frac{1}{3}\tan\theta_1\tan\theta_3}.
\end{equation}
When $\theta_1=\theta_2=\theta_3=\frac{\pi}{3}$, the YBE turns back into the braid relation with $\breve{R}_{\textrm{i}}(\pi/3)=\omega B_{\textrm{i}}$.

In comparison with $d=\sqrt{2}$ for $\breve{R}(\mu)$ related to the Bell basis [3], $\mu=\tan\theta$, and $$\mu_2=\frac{\mu_1+\mu_3}{1+\mu_1\mu_3}.$$

\section{proof of operator multiplication for $u^{\pm},s^{\pm},d^{\pm}$ in the fermionic representation}
In the orthonormal basis of $|r\rangle=r^{\dag}|\textrm{vac}\rangle$, $|g\rangle=g^{\dag}|\textrm{vac}\rangle$ and $|b\rangle=b^{\dag}|\textrm{vac}\rangle$, with the fermionic condition on each site
\begin{equation}
\{x^{\dag}, y\}=\delta_{xy}, \quad \{x^{\dag}, y^{\dag}\}=\{x, y\}=0, \qquad (x, y= r, g, b),
\end{equation}
the operators $u^{\pm},s^{\pm},d^{\pm}$ are expressed as
\begin{equation}\label{}
\begin{split}
&u^{+}=r^{\dag}g, \, s^{+}=\omega g^{\dag}b, \, d^{+}=\omega^2 b^{\dag}r,\\
&u^-=g^{\dag}r, \, s^-=\omega^2 b^{\dag}g, \, d^-=\omega r^{\dag}b.
\end{split}
\end{equation}
Due to the constraint that the total occupation number for the fermions $r, g, b$ is 1 on each site,
\begin{equation}\label{}
r^{\dag}r+g^{\dag}g+b^{\dag}b=1.
\end{equation} 
In the basis of $|r\rangle$, $|g\rangle$ and $|b\rangle$, we have the relation
\begin{equation}
w^{\dag}xy^{\dag}z|n\rangle=\delta_{xy}w^{\dag}z|n\rangle-w^{\dag}z^{\dag}xy|n\rangle=\delta_{xy}w^{\dag}z|n\rangle. \quad(w, x, y, z,n\in\{r, g, b\})
\end{equation} 
i.e. acting two adjoint annihilation operators on the single fermionic occupation basis. Then multiplication of the operators $u^{\pm},s^{\pm},d^{\pm}$ are easily checked in the basis of $|r\rangle$, $|g\rangle$ and $|b\rangle$,
\begin{eqnarray*}
&&[u^+]^2=r^{\dag}gr^{\dag}g=-r^{\dag}r^{\dag}gg=0;\\
&& [u^-]^2=g^{\dag}rg^{\dag}r=-g^{\dag}g^{\dag}rr=0;\\
&&[s^+]^2=\omega^2 g^{\dag}bg^{\dag}b=-\omega^2 g^{\dag}g^{\dag}bb=0;\\
&&[s^-]^2=\omega b^{\dag}gb^{\dag}g=-\omega b^{\dag}b^{\dag}gg=0;\\
&&[d^+]^2=\omega b^{\dag}rb^{\dag}r=-\omega b^{\dag}b^{\dag}rr=0;\\
&&[d^-]^2=\omega^2 r^{\dag}br^{\dag}b=-\omega^2 r^{\dag}r^{\dag}bb=0;\\
&&u^+s^+=[r^{\dag}g][\omega g^{\dag}b]=\omega r^{\dag}(1-g^{\dag}g)b=\omega r^{\dag}b=d^{-};\\
&&s^+d^+=[\omega g^{\dag}b][\omega^2 b^{\dag}r]=g^{\dag}(1-b^{\dag}b)r= g^{\dag}r=u^{-};\\
&&d^+u^+=[\omega^2 b^{\dag}r][r^{\dag}g]=\omega^2 b^{\dag}(1-r^{\dag}r)g=\omega^2 b^{\dag}g=s^{-};\\
&&s^-u^-=[\omega^2 b^{\dag}g][g^{\dag}r]=\omega^2 b^{\dag}(1-g^{\dag}g)r=\omega^2 b^{\dag}r=d^{+};\\
&&d^-s^-=[\omega r^{\dag}b][\omega^2 b^{\dag}g]=r^{\dag}(1-b^{\dag}b)g=r^{\dag}g=u^{+};\\
&&u^-d^-=[g^{\dag}r][\omega r^{\dag}b]=\omega g^{\dag}(1-r^{\dag}r)b=\omega g^{\dag}b=s^{+}.
\end{eqnarray*}
Hence Temperley-Lieb algebraic relation represented by fermions can also be checked.
\section{Derivation of 3-body Hamiltonian}

The 3-body S-matrix constrained by YBE is
\begin{eqnarray}
 \breve{R}_{123}(\theta_1,\theta_2,\theta_3)
&=&\breve{R}_{12}(\theta_1) \breve{R}_{23}(\theta_2) \breve{R}_{12}(\theta_3)\nonumber\\
&=&\breve{R}_{23}(\theta_3) \breve{R}_{12}(\theta_2) \breve{R}_{23}(\theta_1).
 \end{eqnarray}
Regarding $ \breve{R}_{123}$ as the unitary evolution of system, one can construct the 3-body Hamiltonian

\begin{equation}\label{3bodySE}
\hat{H}_{\textrm{123}}(t)=\textrm{i}\hbar\tfrac{\partial\breve{R}_{123}}{\partial t}\breve{R}_{123}^{-1}.
\end{equation}
Here we note that due to the constraint of equation (12), there are two of the three parameters $\theta_1$, $\theta_2$ and $\theta_3$ are free Supposing that $\theta_1$ and $\theta_2$ is time dependent, we obtain 
 
 \begin{widetext}
 \begin{equation}\label{H123}
 \begin{split}
\hat{H}_{\textrm{123}}=&\textrm{i}\hbar\dot{\theta}_1\tfrac{\partial\breve{R}_{12}(\theta_1)}{\partial \theta_1}\breve{R}_{12}^{-1}(\theta_1)+\textrm{i}\hbar\dot{\theta}_2\breve{R}_{12}(\theta_1)\left[\tfrac{\partial\breve{R}_{23}(\theta_2)}{\partial \theta_2}\breve{R}_{23}^{-1}(\theta_2)\right]\breve{R}_{12}^{-1}(\theta_1)\\
=&-\hbar\dot{\theta_1}(\tfrac{3}{\sqrt{2}}T_{1}-1)-\hbar\dot{\theta_2}\breve{R}_{12}(\theta_1)\left[\tfrac{3}{\sqrt{2}}T_{2}-1\right]\breve{R}_{12}^{-1}(\theta_1)\\
=&-\tfrac{8}{9}\hbar\dot{\theta_2}\sin\theta_1\sin(\theta_1+\pi/3)\omega^2(C_{\textrm{1}}^{\dagger}C_{\textrm{3}}+C_{\textrm{1}}C_{\textrm{3}}^{\dagger})+\hbar\dot{\theta_2}(\tfrac{8}{9}\sin^2\theta_1-\tfrac{2}{3})\omega^2(C_{\textrm{2}}^{\dagger}C_{\textrm{3}}+C_{\textrm{2}}C_{\textrm{3}}^{\dagger})\\
&-\tfrac{8}{9}\hbar\dot{\theta_2}\sin\theta_1\sin(\theta_1-\pi/3)(C_{\textrm{1}}^{\dagger}C_{\textrm{2}}^{\dagger}C_{\textrm{3}}^{\dagger}+C_{\textrm{1}}C_{\textrm{2}}C_{\textrm{3}})-\tfrac{2}{3}\hbar\dot{\theta_1}\omega^2(C_{\textrm{1}}^{\dagger}C_{\textrm{2}}+C_{\textrm{1}}C_{\textrm{2}}^{\dagger})+\hbar\tfrac{1}{3}(\dot{\theta_1}+\dot{\theta_2}).
\end{split}
\end{equation}
 \end{widetext}
 Ignoring the constant term, we have
 \begin{eqnarray*}
 &&\alpha=-\tfrac{2}{3}\hbar\dot{\theta_1},\\
 &&\beta=-\tfrac{8}{9}\hbar\dot{\theta_2}\sin\theta_1\sin(\theta_1+\pi/3),\\
 &&\gamma=\hbar\dot{\theta_2}(\tfrac{8}{9}\sin^2\theta_1-\tfrac{2}{3}),\\
 &&\kappa=-\tfrac{8}{9}\hbar\dot{\theta_2}\sin\theta_1\sin(\theta_1-\pi/3),
 \end{eqnarray*}
 for
  \begin{eqnarray}
 \hat{H}_{123}&=&\omega^2[\alpha(C_{\textrm{1}}^{\dagger}C_{\textrm{2}}+C_{\textrm{1}}C_{\textrm{2}}^{\dagger})+\beta(C_{\textrm{1}}^{\dagger}C_{\textrm{3}}+C_{\textrm{1}}C_{\textrm{3}}^{\dagger})\\
 &&+\gamma(C_{\textrm{2}}^{\dagger}C_{\textrm{3}}+C_{\textrm{2}}C_{\textrm{3}}^{\dagger})]+\kappa(C_{\textrm{1}}^{\dagger}C_{\textrm{2}}^{\dagger}C_{\textrm{3}}^{\dagger}+C_{\textrm{1}}C_{\textrm{2}}C_{\textrm{3}}).\nonumber
 \end{eqnarray}
 
 \section{ Symmetry operators of $\hat{H}_{123}$}
 In this section, we show that there are only two independent symmetry operators of 3-body Hamiltonian $\hat{H}_{123}$ . Let us first transform equation (19) into matrix tensor product form under SU(3)  Jordan-Wigner transformation 
 \begin{eqnarray}
&&U_{\textrm{n}}^{\dagger}=\left[\prod_{\textrm{i}=1}^{\textrm{n}-1}[T_{2}^{(1)}]_{\textrm{i}}\right]u^{+}_{n},\quad U_{\textrm{n}}=\left[\prod_{\textrm{i}=1}^{\textrm{n}-1}[T_{3}^{(1)}]_{\textrm{i}}\right]u^{-}_{n};\nonumber\\
&&S_{\textrm{n}}^{\dagger}=\left[\prod_{\textrm{i}=1}^{\textrm{n}-1}[T_{2}^{(1)}]_{\textrm{i}}\right]s^{+}_{n},\quad S_{\textrm{n}}=\left[\prod_{\textrm{i}=1}^{\textrm{n}-1}[T_{3}^{(1)}]_{\textrm{i}}\right]s^{-}_{n};\\
&&D_{\textrm{n}}^{\dagger}=\left[\prod_{\textrm{i}=1}^{\textrm{n}-1}[T_{2}^{(1)}]_{\textrm{i}}\right]d^{+}_{n}, \quad D_{\textrm{n}}=\left[\prod_{\textrm{i}=1}^{\textrm{n}-1}[T_{3}^{(1)}]_{\textrm{i}}\right]d^{-}_{n}.\nonumber
\end{eqnarray}
We have
\begin{eqnarray*}
&&C_{1}^{\dag}=T_{1}^{(2)}\otimes T_{1}^{(1)}, \quad C_{1}=T_{1}^{(3)}\otimes T_{1}^{(1)};\\
&&C_{2}^{\dag}=\omega T_{2}^{(2)}\otimes T_{1}^{(1)}, \, \, C_{2}=T_{3}^{(3)}\otimes T_{1}^{(1)};\\
&&C_{3}^{\dag}=T_{2}^{(1)}\otimes T_{1}^{(2)}, \quad C_{3}=T_{3}^{(1)}\otimes T_{1}^{(3)};\\
&&C_{4}^{\dag}=\omega T_{2}^{(1)}\otimes T_{2}^{(2)}, \, \, C_{4}=T_{3}^{(1)}\otimes T_{3}^{(3)}.
\end{eqnarray*}
Then the Hamiltonian reads
\begin{eqnarray*}
\hat{H}_{123}&=&\alpha\left(\omega^2T_{2}^{(1)}\otimes T_{1}^{(1)}+\omega T_{3}^{(1)}\otimes T_{1}^{(1)}\right)+\beta\left(\omega^2T_{3}^{(2)}\otimes T_{1}^{(3)}+\omega^2 T_{2}^{(3)}\otimes T_{1}^{(2)}\right)\\
&&+\gamma\left(\omega^2T_{1}^{(2)}\otimes T_{1}^{(3)}+\omega T_{1}^{(3)}\otimes T_{1}^{(2)}\right)+\kappa\left(\omega T_{3}^{(3)}\otimes T_{1}^{(2)}+T_{2}^{(2)}\otimes T_{1}^{(3)}\right).
\end{eqnarray*}
Based on the the algebraic relation in equation (1), we try to find the independent symmetry operators without linear composition. The general form of the operators can be expressed as
\begin{equation}
\Gamma_{\textrm{i}}=T_{a}^{(x)}\otimes T_{b}^{(y)}.
\end{equation}
Here $a$, $b$, $x$ and $y$ are to be determined, $\{a, b, x, y\}\in\{1, 2, 3\}$. The condition is that each term in $\hat{H}_{123}$ commutes with $\Gamma_{\textrm{i}}$,
\begin{eqnarray*}
&&[T_{2}^{(1)}\otimes T_{1}^{(1)}, T_{a}^{(x)}\otimes T_{b}^{(y)}]=0;\\
&&[T_{3}^{(2)}\otimes T_{1}^{(3)}, T_{a}^{(x)}\otimes T_{b}^{(y)}]=0;\\
&&[T_{1}^{(2)}\otimes T_{1}^{(3)}, T_{a}^{(x)}\otimes T_{b}^{(y)}]=0;\\
&&[T_{2}^{(3)}\otimes T_{1}^{(2)}, T_{a}^{(x)}\otimes T_{b}^{(y)}]=0.
\end{eqnarray*}
After direct calculation, one obtains
\begin{equation}
x=1, \, a=b\in\{1,2,3\}, \, y\in\{1,2,3\}.
\end{equation}
Totally, there are 8 cases, 
\begin{eqnarray*}
&&\Gamma_{\textrm{1}}=T_{1}^{(1)}\otimes T_{1}^{(2)}, \Gamma_{\textrm{5}}=T_{1}^{(1)}\otimes T_{1}^{(3)};\\
&&\Gamma_{\textrm{2}}=T_{2}^{(1)}\otimes T_{2}^{(1)}, \Gamma_{\textrm{6}}=T_{3}^{(1)}\otimes T_{3}^{(1)};\\
&&\Gamma_{\textrm{3}}=T_{2}^{(1)}\otimes T_{2}^{(2)}, \Gamma_{\textrm{7}}=T_{3}^{(1)}\otimes T_{3}^{(3)};\\
&&\Gamma_{\textrm{4}}=T_{2}^{(1)}\otimes T_{2}^{(3)}, \Gamma_{\textrm{8}}=T_{3}^{(1)}\otimes T_{3}^{(2)}.
\end{eqnarray*}
But it is easy to check that 
\begin{eqnarray*}
&&\Gamma_{\textrm{5}}\propto[\Gamma_{\textrm{1}}]^2,\quad \Gamma_{\textrm{6}}\propto[\Gamma_{\textrm{2}}]^2,\\
&&\Gamma_{\textrm{7}}\propto[\Gamma_{\textrm{3}}]^2,\quad \Gamma_{\textrm{8}}\propto[\Gamma_{\textrm{4}}]^2,\\
&&\Gamma_{\textrm{3}}=\Gamma_{\textrm{2}}\Gamma_{\textrm{1}},\,\Gamma_{\textrm{4}}=\Gamma_{\textrm{2}}[\Gamma_{\textrm{1}}]^2.
\end{eqnarray*}
Then only 2 independent operators are left,
\begin{eqnarray}
&&\Gamma_{\textrm{1}}=T_{1}^{(1)}\otimes T_{1}^{(2)},\\
&&\Gamma_{\textrm{2}}=T_{2}^{(1)}\otimes T_{2}^{(1)}.
\end{eqnarray}  
Making inverse SU(3) J-W transformation, we can redefine that
\begin{eqnarray}
 &&P=\Gamma_{\textrm{2}}=C_1C_2^{\dag}C_3C_4^{\dag},\\
 &&\Gamma=\Gamma_{\textrm{1}}=\omega C_1^{\dag}C_2C_3^{\dag}.
\end{eqnarray} 
Here we define $P$ as the $\mathbb{Z}_3$ $\omega$-parity operator. Hence we find the two symmetry operators of $\hat{H}_{123}$.   

\bibliographystyle{unsrt}